\documentclass[aps, twocolumn, 10pt, showpacs, longbibliography, superscriptaddress]{revtex4-1}
\usepackage{graphicx}
\usepackage{lipsum}
\usepackage{dcolumn}
\usepackage{bm}
\usepackage{color}
\usepackage{amsmath, amsthm, amssymb}
\usepackage[modulo]{lineno}
\usepackage[usenames,dvipsnames]{xcolor}
\usepackage[normalem]{ulem}
\usepackage{comment}
\newcommand{\ba}{\begin{eqnarray}}
\newcommand{\ea}{\end{eqnarray}}

\usepackage[colorlinks=true, urlcolor=violet, linkcolor=blue, citecolor=red, hyperindex=true, linktocpage=true]{hyperref}

\usepackage{float}
\floatstyle{boxed}
\graphicspath{ {Figures/} }

\begin{document}
\title{Graphene calorimetric single-photon detector}

\author{Bevin Huang}%
\altaffiliation{These authors contributed equally to this work}
\affiliation{Intelligence Community Postdoctoral Research Fellowship Program, Massachusetts Institute of Technology, Cambridge, MA 02139}
\author{Ethan G. Arnault}%
\altaffiliation{These authors contributed equally to this work}
\affiliation{Department of Electrical Engineering and Computer Science, Massachusetts Institute of Technology, Cambridge, MA 02139}
\author{Woochan Jung}%
\altaffiliation{These authors contributed equally to this work}
\affiliation{Department of Physics, Pohang University of Science and Technology, Pohang 790-784, Republic of Korea}

\author{Caleb Fried}%
\affiliation{Department of Electrical Engineering and Computer Science, Massachusetts Institute of Technology, Cambridge, MA 02139}

\author{B. Jordan Russell}%
\affiliation{Department of Physics, Washington University in St.\ Louis, St.\ Louis, MO, USA}

\author{Kenji Watanabe}
\affiliation{Research Center for Electronic and Optical Materials, National Institute for Materials Science, 1-1 Namiki, Tsukuba 305-0044, Japan}
\author{Takashi Taniguchi}
\affiliation{Research Center for Materials Nanoarchitectonics, National Institute for Materials Science,  1-1 Namiki, Tsukuba 305-0044, Japan}

\author{Erik A. Henriksen}%
\affiliation{Department of Physics, Washington University in St.\ Louis, St.\ Louis, MO, USA}

\author{Dirk Englund}%
\affiliation{Department of Electrical Engineering and Computer Science, Massachusetts Institute of Technology, Cambridge, MA 02139}
\author{Gil-Ho Lee}%
\email{lghman@postech.ac.kr}
\affiliation{Department of Physics, Pohang University of Science and Technology, Pohang 790-784, Republic of Korea}
\author{Kin Chung Fong}%
\altaffiliation{Present address: k.fong@northeastern.edu, Northeastern University}
\affiliation{RTX BBN Technologies, Quantum Engineering and Computing Group, Cambridge, Massachusetts 02138, USA}

\date{\today}
\begin{abstract} Single photon detectors (SPDs) \cite{Hadfield.2009,Eisaman.2011} are essential technology in quantum science, quantum network, biology, and advanced imaging \cite{Couteau.2023,Bruschini.Charbon.2019,Kirmani.2014}. To detect the small quantum of energy carried in a photon, conventional SPDs rely on energy excitation across either a semiconductor bandgap or superconducting gap. While the energy gap suppresses the false-positive error, it also sets an energy scale that can limit the detection efficiency of lower energy photons and spectral bandwidth of the SPD \cite{Ceccarelli.Osellame.2021,Seifert.2020,Verma.2021}. Here, we demonstrate an orthogonal approach to detect single near-infrared photons using graphene calorimeters \cite{Fong.2012,Du.2014,Walsh.2017,Lee.2020,Kokkoniemi.2020}. By exploiting the extremely low heat capacity of the pseudo-relativistic electrons in graphene near its charge neutrality point \cite{Sarma.2011}, we observe an electron temperature rise up to $\sim$2~K using a hybrid Josephson junction. In this proof-of-principle experiment, we achieve an intrinsic quantum efficiency of 87\% (75\%) with dark count $<$ 1 per second (per hour) at operation temperatures as high as 1.2 K. Our results highlight the potential of electron calorimetric SPDs for detecting lower-energy photons from the mid-IR to microwave regimes, opening pathways to study space science in far-infrared regime \cite{Echternach.2018,day.2024}, to search for dark matter axions \cite{Hochberg.2019,Dixit.2021}, and to advance quantum technologies across a broader electromagnetic spectrum \cite{Lauk.2020}.\end{abstract}
\maketitle

Photons are the quantum particles of electromagnetic field, each carrying a small amount of energy. This makes their detection, particularly at lower energies, challenging. Many conventional single-photon detectors (SPDs) operate by a photo-excitation across an energy gap. In semiconductor-based avalanche photodiodes \cite{Ceccarelli.Osellame.2021}, the excitation creates an electron-hole pair across the bandgap. In superconducting nanowires \cite{Gol'tsman.2001,Charaev.2023,Verma.2021} or kinetic inductance detectors \cite{day.2024}, the excitation breaks Cooper pairs and promotes quasiparticles above the superconducting gap, $\Delta_s$. In each case, the energy gap provides a mechanism to distinguish photons from dark counts caused by fluctuations, but also limits the detection of lower-energy photons. Detecting single-photons by directly sensing their energy (e.g. transition-edge sensors \cite{Ullom.Bennett.2015}) can potentially resolve this dilemma. However, these SPDs typically have a substantial heat capacity, limiting their efficacy in detecting lower-energy photons.

\begin{figure*}[t]
\includegraphics[width=5in]{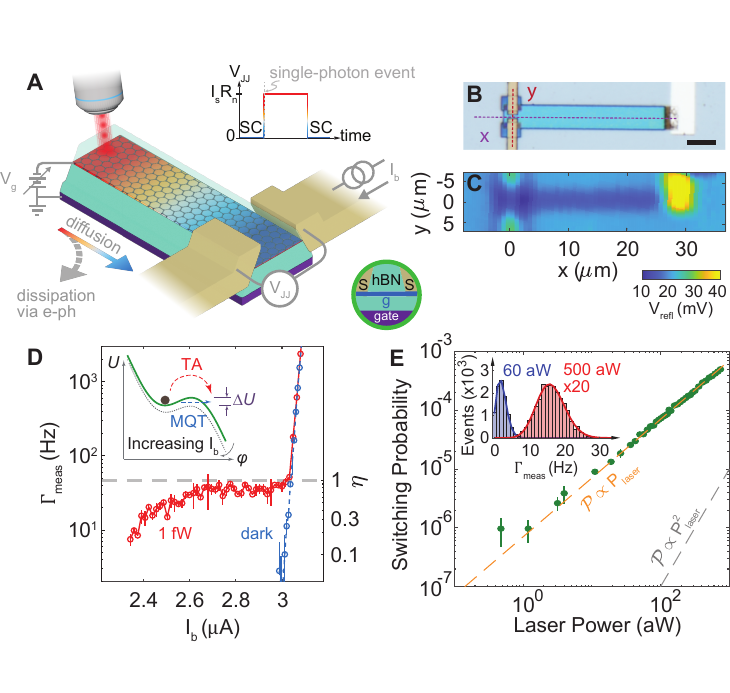}
\caption{\textbf{Graphene calorimetric single-photon detector. (A)} Illustration of the experiment. A photon is absorbed on one end of the graphene, heating the electrons. The hot electrons then diffuse throughout the graphene while dissipating into the lattice via electron-phonon coupling. Inset: Junction voltage $V_{\rm{JJ}}$ as a function of time. When the photon is absorbed, the junction switches from superconducting to resistive causing a voltage drop across the junction. The device is then reset to the superconducting state.  \textbf{(B)} Optical image of one of the graphene calorimetric SPDs. Scale bar is 5~$\mu$m. \textbf{(C)} The 2D reflectometry measurement of the device at low temperature. \textbf{(D)} Switching rate $\Gamma_{\text{meas}}$ vs. current bias $I_b$ for laser on (red) and off (blue). The mean value of quantum efficiency, $\eta$, in the plateau from 2.7 $\mu$A $< I_b <$ 3 $\mu$A with 1 fW of laser power is 0.77 $\pm$ 0.08. Inset: Current-biased JJ can be described as a macroscopic quantum phase particle subjected to a tilted-washboard potential in the Resistively Capacitance Shunted Junction model. When dark, the junction is nominally in the MQT regime, however the photon raises the temperature of the junction causing a thermally activated switching event. \textbf{(E)} Junction switching probability vs. laser power at $V_{\text{gate}} = 2$~V and $I_b/\langle I_s \rangle \simeq$ 0.87. The junction switching probability is linearly proportional to the laser power, confirming a single photon can switch the Josephson junction from superconducting to resistive. Deviation from the linear trend at lower powers is due to dark counts. Inset: Histogram of switching events with 60 aW (blue) and 500 aW (red) of laser powers adhere to Poissonian statistics.}
\label{fig:Fig1}
\end{figure*}

Graphene presents a promising material for a calorimetric SPDs \cite{Fong.2012,Du.2014,Walsh.2017,Lee.2020,Kokkoniemi.2020}. Specifically, graphene electrons' vanishing density of states near the charge neutrality point result in a very low electronic specific heat ($\sim$1~$k_B/\mu$m$^2$ with $k_B$ being Boltzmann constant) and suppressed electron-phonon (E-Ph) coupling \cite{Hwang.2008,song.levitov.2015}. Consequently, the energy deposited by a single photon is confined to the graphene electrons, leading to an exceptionally large rise in graphene electron temperature, $T_e$, for calorimetric SPD. Yet, utilizing graphene electrons as a single-photon calorimeter has its own challenges. For instance, the fleeting $T_e$ rise requires simultaneously fast and accurate readouts to measure photon absorption. Moreover, infrared photons may interact directly with the superconducting electrodes, generating quasiparticles that interfere with the operation of the electron calorimeter. Despite remarkable progress in achieving graphene bolometers with sensitivities at the fundamental thermodynamic limit \cite{Lee.2020,Kokkoniemi.2020}, a graphene calorimetric SPD remains elusive. In this work, we implement an optical scanner at cryogenic temperatures. We demonstrate that, upon absorption, the internal energy from a single photon can heat up the electrons and propagate through the graphene, thermally triggering the switching of a Josephson junction \cite{Walsh.2017,Zgirski.2018}. Our experiment allows us to achieve high quantum efficiency, low dark count SPDs based on the calorimetric effects in pseudo-relativistic electrons in graphene.

Fig. 1A depicts our setup in a dilution refrigerator at temperatures $T_0 \simeq 20$ mK. 1550-nm light is routed through a single-mode optical fiber into a collimator and subsequently focused by an aspherical lens. This optical set-up is affixed on top of a three-axis piezoelectric stage which can steer the highly attenuated laser source from room temperature to the graphene absorber of area 4 $\mu$m $\times$ 25 $ \mu$m (Fig. 1B) with sub-$\mu$m spatial precision and a beam spot size of 4 $\mu$m. We can scan over the device and measure the laser reflectometry signal, $V_{\text{refl}}$, (Methods) to identify features on the chip and ensure the location of our beam spot over the graphene (Fig. 1C). Upon absorption, the single photon will create a hotspot of heated electrons \cite{Tielrooij.2013}, which will quickly diffuse across the graphene \cite{Block.2021}, dissipating energy to the graphene lattice via E-Ph coupling. When diffusion dominates over E-Ph dissipation, the entire graphene area reaches a uniform $T_e$ that peaks at $T_{1p} = \sqrt{2h\nu/\gamma_{\text{S}}\mathcal{A}+T_0^2}$ \cite{Walsh.2017} with $h$ being the Planck constant, $\nu$ photon frequency, $\mathcal{A}$ the graphene area, and $\gamma_{\text{S}}$ Sommerfeld constant, which is the ratio of the electronic specific heat per unit area, $c_e$, to $T_e$. For the graphene in Fig.~1B, $T_{1p} \sim$2~K. To overcome the challenge of measuring the rise of $T_e$ in a short time scale of a few tens of ns \cite{Lee.2020}, we use a graphene-based Josephson junction (GJJ) (fabrication details in Methods), whose response rate is on the order of the plasma frequency \cite{Zgirski.2018}, $\omega_p \gtrsim$ 100 GHz, for the SPD readout.

We can use the Resistively and Capacitively Shunted Junction (RCSJ) model \cite{Tinkham} to understand how a single photon switches the GJJ. In RCSJ, a macroscopic quantum phase particle with a phase difference, $\varphi$, between the two superconducting electrodes is subject to a washboard potential (Fig. 1D inset). When the phase particle is trapped initially in a local minima, i.e. $d\varphi/dt = 0$, the voltage drop across the GJJ is zero. The bias current, $I_b$, running through the GJJ tilts the washboard potential and the phase particle stochastically escapes from the minimum. When it escapes, either by thermal activation (TA) \cite{Martinis.1987} over or macroscopic quantum tunneling (MQT) \cite{Devoret.1985} through the barrier, $\Delta U$, the voltage drop across the GJJ becomes finite and the GJJ switches to the normal resistive state at a switching current $I_s$ (Fig. 1A inset). When the phase particle is retrapped at a retrapping current $I_r$, the GJJ switches back to the supercurrent state. The hysteretic behavior, i.e. $I_s>I_r$, frequently observed in graphene-based GJJs due to self-Joule heating \cite{Courtois.2008,Borzenets.2016nj,Lee.2020}, is useful to our investigation. When the GJJ latches into the resistive state after switching, we register a click, reset the bias current, and over time, measure the switching statistics \cite{Walsh.2021} under different light intensities, densities of graphene electrons, and temperatures. 

Fig. 1D shows the measured switching rate, $\Gamma_{\text{meas}}$, versus $I_b$. We designate  $\Gamma_{\text{meas}}$ without photons as the dark count rate, $\Gamma_{\text{dark}}$, which is governed by quantum fluctuations. The fit (dashed line) appears nearly straight in the log-linear plot because the rate of MQT follows activation theory, i.e. $\propto\exp{(-7.2\Delta U/\hbar \omega_p)}$ \cite{Devoret.1985}, where $\hbar$ is the reduced Planck constant and $\hbar \omega_p$ is the zero-point fluctuation that assists the GJJ phase particle tunneling through the potential barrier $\Delta U$.

With 1-fW illumination, $\Gamma_{\text{meas}}$ is considerably higher than in the dark. At a constant photon flux, $\Gamma_{\text{meas}}$ increases monotonically with $I_b$. The absorbed photons can switch the GJJ more readily at a higher $I_b$ because the phase particle can escape over a lower $\Delta U$. Below $\sim$2.7~$\mu$A, the junction may retrap before detection. This leads to false negative counts resulting in a reduction of quantum efficiency, $\eta$, defined as the number of measured single photons over the total number of photons absorbed into the detector. At $\sim$2.7~$\mu$A, $\Gamma_{\text{meas}}$ starts to saturate, signifying that $\eta$ is approaching near unity, similar to superconducting nanowire detectors \cite{Verma.2021}. When $I_b > 3$ $\mu$A, the GJJ switches spontaneously by the MQT mechanism such that $\Gamma_{\text{meas}}$ is dominated by junction self-switching. The nonlinear $\Gamma_{\text{meas}}$ in the log-linear plot deviates from activation theory and underscores the detection of single photons as discrete events rather than a continuous heating \cite{Walsh.2021}. Calibrated using $V_{\text{refl}}$ \cite{si}, 1-fW photon illumination corresponds to 45 photons/s absorbed into the graphene. The right $y$-axis in Fig. 1D shows that $\eta\simeq 0.77\pm0.08$ when $\Gamma_{\text{meas}}$ saturates.

We can prove that each of the GJJ switching events is triggered by a single photon \cite{Gol'tsman.2001}. For a coherent state, the probability of an $m$ photon state, $\mathcal{P}_c(m)$, with a mean photon number, $\mu$, follows the Poisson distribution, i.e. $e^{-\mu} \mu^m/m!$. When $\mu\ll 1$, $\mathcal{P}_c(m=1)$ grows linearly with $\mu$, and hence the laser power. We measure the switching events of our detector over a range of laser powers for 300 seconds to obtain the switching rate, $\Gamma_{\text{meas}}$. We then calculate switching probability, $\mathcal{P} =\Gamma_{\text{meas}}/\mathcal{B}$, where $\mathcal{B}$ is the bandwidth of our detector upper bounded by the lowpass filters (30 kHz) used in biasing and measuring the GJJ \cite{Gol'tsman.2001,Walsh.2021}. Fig. 1E shows that $\mathcal{P}$ depends linearly on laser power over several orders of magnitude, proving our detector is single-photon sensitive. Furthermore, Fig. 1E inset plots the distribution of $\Gamma_{\text{meas}}$. The histogram follows the Poisson statistics (solid lines) and the standard deviation constitutes the shot noise of uncorrelated photons from the coherent source.

\begin{figure*}
\includegraphics[width=5in]{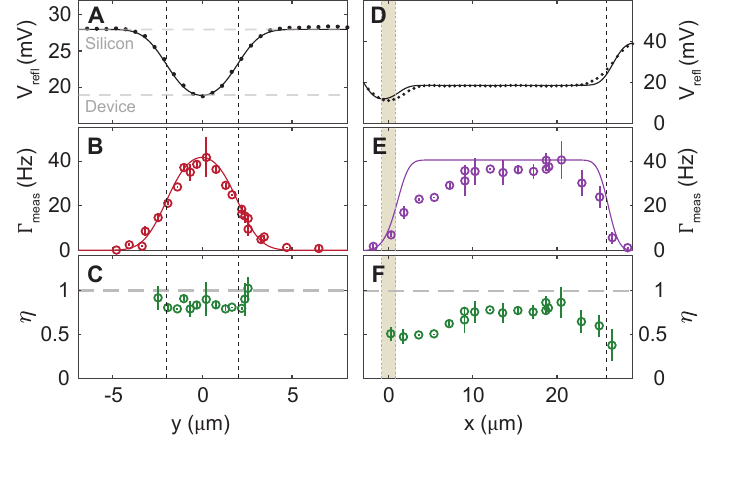}
\caption{\textbf{Scanning the beam spot across the graphene. (A-C)} Scanning in the transverse ($y$-) direction marked by the maroon dashed line in Fig.~1B. Vertical dashed lines mark the graphene location. \textbf{(D-F)} Scanning in the longitudinal ($x$-) direction. Vertical light yellow box designates the Josephson-junction location. Vertical dashed line marks the graphene location. \textbf{(A, D)} Reflectance signal (dots) and fitting to the convolution integral (solid line). The dashed gray lines indicate the calculated reflectance values of silicon and the graphene heterostructure. \textbf{(B, E)} The measured and expected switching rate (open dots and solid line, respectively) based on the convolution between graphene and a Gaussian beam of spot size 4 $\mu m$. \textbf{(C, F)} Quantum efficiency calculated by dividing $\Gamma_{\rm{meas}}$ with the absorbed photon rate. The mean value of $\eta$ in \textbf{(C)} is 0.85 $\pm$ 0.07. All data were taken at $I_b/\langle I_s \rangle \simeq$ 0.87 and $V_{\text{gate}} = 2$~V.}
\label{fig:Fig2}
\end{figure*}

To demonstrate the calorimetric effect, we study the photon absorption by measuring $\Gamma_{\text{meas}}$ as the laser scans across the graphene. Fig. 2A shows $V_{\text{refl}}$ as the beam rasters the transverse ($y$-) axis, 10 $ \mu$m away from the GJJ and parallel to the red dashed line in Fig. 1B. The data agrees well with the calculated spatial dependence of the $V_{\text{refl}}$ (solid line) by convolving a Gaussian profile of the 4-$\mu$m beam spot with a boxcar function representing the spatial extent of the graphene heterostructure (marked by the vertical dashed lines) \cite{si}. Specifically, the measured $V_{\text{refl}}$ also matches to our calculated ratio of the reflectance (horizontal dashed lines) of silicon to that of the graphene heterostructure. The excellent agreement supports that the calibration of the photon absorption into the monolayer graphene due to the interference effect from the graphene heterostructure is about 0.61\% \cite{si}. This can be improved up to 99\% by a photonic cavity \cite{Gan.2012,Furchi.Mueller.2012,Vasic.2014}.

Fig. 2B plots $\Gamma_{\text{meas}}(y)$ which resembles $V_{\text{refl}}(y)$, indicating that the absorbed photon switches the GJJ. When the beam spot is completely off the graphene, we measured zero $\Gamma_{\text{meas}}(|y| \geq 5~\mu \text{m})$, confirming that the stray light does not contribute to the measured single-photon counts. We normalize $\Gamma_{\text{meas}}(y)$ by the expected rate of absorbed photon to estimate $\eta$ \cite{si} (Fig. 2C). Contrary to the variations of $\Gamma_{\text{meas}}(y)$ and $V_{\text{refl}}(y)$, $\eta(y)$ remains roughly a constant with an average value of $\sim$0.8 when the beam spot illuminates the graphene.

To investigate the heat propagation in the calorimetric SPD, we measured $\Gamma_{\text{meas}}$ in the longitudinal ($x$-) direction of the device (Fig.\ 1B purple dashed line). Fig. 2D and E plots $V_{\text{refl}}(x)$ and $\Gamma_{\text{meas}}(x)$, respectively. By positioning the beam spot far away from the GJJ, we can ensure no clicks are due to Cooper pair breaking from photon exposure in the superconducting electrodes.  Similar to $\Gamma_{\text{meas}}(y)$, $\Gamma_{\text{meas}}(x)$ subsides when the beam spot moves off the graphene absorber. Interestingly, $\Gamma_{\text{meas}}(x)$ remains high when the beam spot is positioned far away from the GJJ. By approximating our long flake as one-dimensional, we can understand this behavior using a dissipative diffusion equation \cite{Fried.Fong.2024}: \ba \frac{\partial}{\partial t} T_e^2= \mathcal{D}\frac{\partial^2}{\partial x^2}T_e^2 - \frac{1}{\tau_{\text{ep}}}\left( T_{e}^{\delta} - T_{0}^{\delta} \right) \label{eqn:heatdiffeqn2}\ea with $\tau_{\text{ep}}$ being the decaying time constant of the E-Ph dissipation, $\delta$ being the E-Ph coupling power law, and $\mathcal{D}$ being the electronic diffusion constant which is given by $\sigma\mathcal{L}_0/\gamma_{\text{S}}$ where $\sigma$ and $\mathcal{L}_0$ are the electrical conductivity and Lorenz number, respectively. The first and second term on the right-hand side of Eqn. \ref{eqn:heatdiffeqn2} represent the heat diffusion and dissipation, respectively. The ratio of these coefficients determines the characteristic length scale of heat diffusion, $l_{\text{D}} = \sqrt{\mathcal{D}\tau_{\text{ep}}} \simeq 230~\mu$m \cite{si}, which is much longer than our sample length, leading to a small variation in $\Gamma_{\text{meas}}(x)$.

\begin{figure*}
\includegraphics[width=6in]{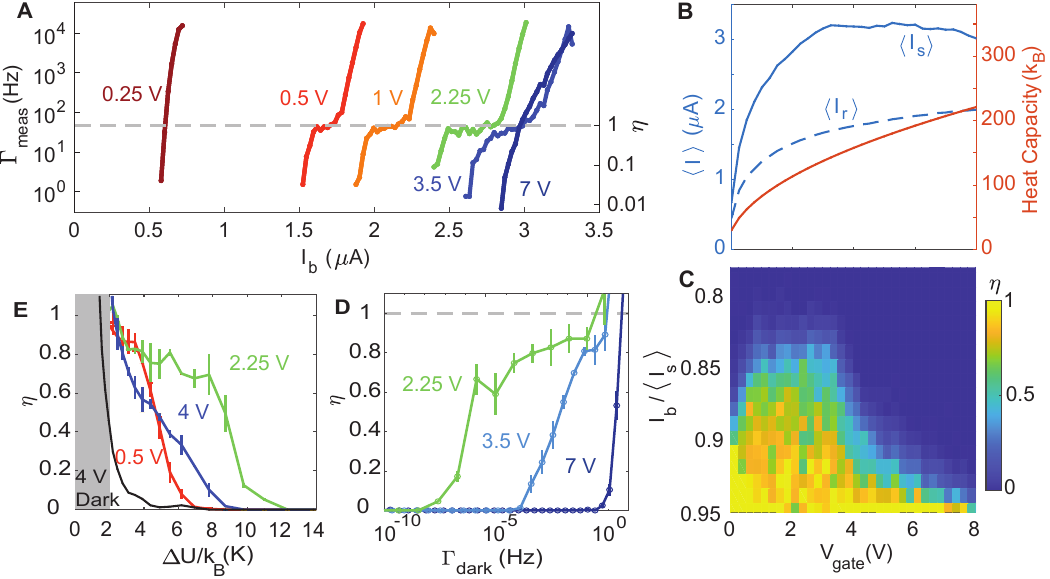}
\caption{\textbf{Dependence of calorimetric SPD performance on electron density. (A)} Gate dependence of $\Gamma_{\text{meas}}$ and $\eta$ under 1 fW of laser power with the beam spot at position $(x, y) = (6, 0)$ $\mu$m. Charge neutrality is at $V_{\text{gate}} = -0.15$ V. \textbf{(B)} Gate dependence of measured $\langle I_s\rangle$ and $\langle I_r\rangle$ (blue), and calculated heat capacity of the graphene absorber (orange). \textbf{(C)} $\eta$ as a function of $V_{\text{gate}}$ and $I_b$. The region of high $\eta$ (yellow) under a relatively small $I_b$ marks the optimal performance of the SPD. \textbf{(D)} Tradespace between $\eta$ vs. dark count ($\Gamma_{\text{dark}}$) for three different $V_{\text{gate}}$. At an optimal $V_{gate}$ of 2.25 V, $\eta \approx$ 0.87 (0.75) for a $\Gamma_{\text{dark}}$ on the order of 1 photon/s (1~photon/hour). (\textbf{E}) $\eta$ vs. $\Delta U/k_B$ of the washboard potential for three different $V_{\text{gate}}$. Gray box indicates the region where dominated by self-switching of the GJJ. At $V_{\text{gate}} =$ 2.25 V, a single photon can induce the escape of the GJJ phase particle from a $\Delta U /k_B$ of $\sim$8 K.} 
\label{fig:Fig3}
\end{figure*}

Fig. 2F plots the $\eta(x)$. The suppression near both ends of the graphene are potentially due to the scattering of light by the metallic electrodes or, when the beam spot is near the GJJ, heat leakage directly into the superconductors when $k_BT_e$ exceeds $\Delta_s$ ($\sim$1.3 meV for our MoRe electrodes), or when the beam spot is far from the GJJ, due to E-Ph dissipation \cite{Fried.Fong.2024}. After accounting for the reduced area of graphene at the GJJ, $\eta(x)$ exhibits no noticeable variation as the beam spot approaches the GJJ. This suggests the GJJ switching mechanism is primarily governed by the calorimetric effect \cite{Walsh.2017,Zgirski.2018}, rather than quasiparticles \cite{Walsh.2021,Aumentado.2004} generated from the breaking of Cooper pairs when the superconducting electrodes of the GJJ are directly under photon illumination.

The performance of the graphene calorimetric SPD depends on the electron density, $n_e$. Fig. 3A shows $\Gamma_{\text{meas}}$ and $\eta$ vs. $I_b$ at various gate voltages, $V_{\text{gate}}$, with an absorbed photon rate of 45 Hz. As $V_{\text{gate}}$ decreases, $\Gamma_{\text{meas}}$ appears at lower $I_b$ because the GJJ critical current, $I_c$, is determined by $I_c R_n \propto \Delta_s$ \cite{Tinkham}, where $R_n$ is the GJJ normal resistance. As shown in Fig. 3B, when $V_{\text{gate}}$ approaches the charge neutrality point at $-0.15$ V, the number of conduction channels decrease, and hence $I_s$, as a proxy for $I_c$, quenches with increasing $R_n$. The decreasing $I_s$ can degrade the GJJ sensing in two ways: firstly, the reduced Josephson energy makes the GJJ susceptible to thermal noise, pushing the device from the MQT to TA regimes \cite{Lee.2011}; secondly, a smaller $\langle I_s\rangle-I_r$ value encourages the phase particle to retrap without the GJJ latching to the normal state. At $V_{\text{gate}} = 0.25$~V, $\Gamma_{\text{meas}}$ does not rise above the $\Gamma_{\text{dark}}$.

As $V_{\text{gate}}$ increases from 0.25 to $\sim$4 V, $\Gamma_{\text{meas}}$ develops a plateau region near 45 Hz, regardless of $V_{\text{gate}}$ but corresponding to $\eta \geq 0.8$ over $I_b$ ranges $\sim$15\% of $\langle I_s\rangle$, before the steep rise at the high $I_b$. This $\Gamma_{\text{meas}}$ plateau is the saturation of photon counting with a high $\eta$ shown in Fig. 1D. However, when $V_{\text{gate}}$ increases up to 7 V, $\Gamma_{\text{meas}}$ overlaps with $\Gamma_{\text{dark}}$ again. To better observe the performance of the SPD, we compare $\eta$ by normalizing $I_b$ to $\langle I_s\rangle$ at various $V_{\text{gate}}$. Fig. 3C shows the evolution of the plateau and the optimal $V_{\text{gate}}$ (= 2.25 V) where the SPD enjoys simultaneously a high $\eta$ and low $\Gamma_{\text{dark}}$. For $V_{\text{gate}} <$ 0.5~V, the sharp suppression of $\langle I_s\rangle$ leads to a poor $\eta$. For $V_{\text{gate}} \gtrsim$ 2.5~V, $\langle I_s\rangle$ remains roughly a constant. Partially, we attribute the weakening of single-photon detection at higher $V_{\text{gate}}$ to the lower $T_{1p}$ due to a larger $c_e$ at higher $n_e(V_{\rm{gate}})$. However, heat diffusion and thermal decay affect $\eta$ equally for all $V_{\text{gate}}$; $\sigma$, $\gamma_{S}$, and E-Ph coupling scale as $\sqrt{n_e}$, so the $n_e$ dependence cancels out in both $\mathcal{D}$ and $\tau_{\text{ep}}$ \cite{Fried.Fong.2024}. In addition to the calorimetric effect at high $V_{\text{gate}}$, we observe a curve in the log-$\Gamma_{\text{dark}}$ vs. $I_b$ plot that deviates from MQT or TA theory. This indicates additional noise inducing GJJ switching \cite{Fulton.1974}. Better filtering and GJJ sensor design will prevent extra noise from eroding $\eta$ at high $V_{\text{gate}}$.

\begin{figure*}
\includegraphics[width=4.5in]{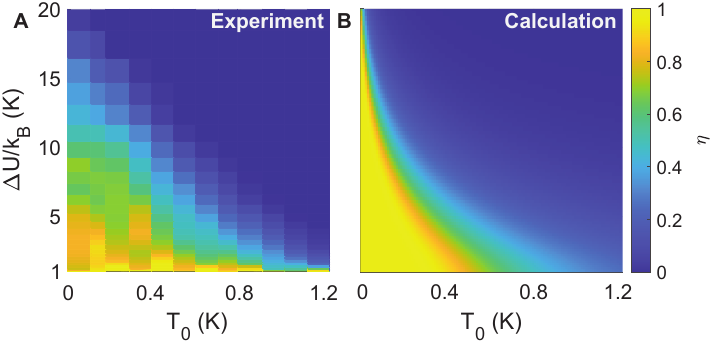}
\caption{\textbf{Intrinsic quantum efficiency as a function of temperature and barrier height of the washboard potential. (A)} Experimental data. \textbf{(B)} Calculation using the temperature dependence of both electronic specific heat in graphene and electron-phonon thermal decay time. Their qualitative agreement suggests a simple calorimetric model in graphene for describing our SPD.}
\label{fig:Fig4}
\end{figure*}

We can benchmark our calorimetric SPD by exploring the competing tradespace between $\eta$ and $\Gamma_{\text{dark}}$ \cite{Walsh.2017}. At higher $I_b$, the GJJ can switch not only by the heat of a single photon, but also spontaneously by thermal or quantum fluctuations. Lowering $\Delta U$ with a higher $I_b$ can improve $\eta$, but at the cost of higher $\Gamma_{\text{dark}}$. Fig. 3D plots the tradespace by extrapolating $\Gamma_{\text{dark}}$ from the MQT that is proven to dominate $\Gamma_{\text{meas}}$ in the absence of photons (Fig. 1D). $\eta$ grows with $\Gamma_{\text{dark}}$ as expected. At $V_{\text{gate}} = 2.25$ V, the device reaches $\eta \approx 0.87$~(0.75) with $\Gamma_{\text{dark}}$ of $\sim$1~photon/s ($\sim$1~photon/hour). In the future, a kinetic inductance readout \cite{Giazotto.2008,Katti.2022} can improve $\eta$ and increase the detector bandwidth, $\mathcal{B}$, while suppressing $\Gamma_{\text{dark}}$.

We can approximate the temperature rise of graphene electrons by a single photon through the $\eta$ dependence on $I_b$. Since the thermal energy from a single photon needs to overcome $\Delta U$ to induce the escape of the phase particle, we obtain $\Delta U/k_B$ as a function of $I_b/\langle I_s\rangle$ and replot cuts from Fig. 3C in Fig. 3E. At $V_{\text{gate}} = 2.25$ V, the data suggests that a single photon can provide enough energy to overcome a $\Delta U/k_B$ of $\sim$8 K, compatible with our estimation of $T_{1p}$ of $\sim$2~K.

To gain more insight into the calorimetric effect, we study $\eta$ versus $\Delta U/k_B$ at various $T_0$ and $V_{\text{gate}} = 2$ V. As shown in Fig. 4A, we are able to detect single photons up to 1.2 K, with a reduced $\eta$ of 0.5. When $T_0$ rises, $\eta$ reduces and the performance of our calorimetric SPD degrades by several mechanisms: (1) the GJJ is subjected to more thermal noise, (2) the rise of $T_e$ from a single photon, $T_{1p}$, diminishes as $c_e$ increases, and (3) $\tau_{\text{ep}}$ shortens with a stronger E-Ph coupling. Between $T_0 =$ 0.02 and 1.2 K, the Josephson plasma frequency remains much greater than $k_BT_0$ because $\langle I_s\rangle$ diminishes only by $\sim$30\%. Therefore, we neglect the temperature dependence of the GJJ and include only the calorimetric effect in graphene, i.e. $c_e \propto T_0$ and $\tau_{\text{ep}}\propto T_0^{2-\delta}$, to model $\eta(T_0)$.

The single-photon enhanced escape probability of the phase particle out of $\Delta U$ can be approximated as $\eta(T_0) = 1-\exp{(-\Gamma_{1p}\tau_{\text{ep}})}$ where $\Gamma_{1p}$ is the enhanced escape rate induced by a single photon that is proportional to $\exp{(-\Delta U/k_BT_{1p})}$ based on the activation theory of a thermal excitation $k_BT_{1p}$. Fig.~4B plots the modeling result using $\delta = 4$ (E-Ph coupling in clean graphene \cite{si}), at $T_{0} = 20$~mK. We find that $\tau_{\text{ep}} = 75$~ns, and $T_{1p}=2.5$~K best matches with the data in Fig. \ref{fig:Fig4}A. Therefore, the three independently evaluated $T_{1p}$ from theory \cite{Walsh.2017,Fried.Fong.2024}, measured $\Delta U$ (Fig. 3E), and thermal modeling (Fig. 4) are mutually consistent. The overall qualitative agreement between the spatial, electronic and thermal dependencies shown in this work demonstrates that a calorimetric model of graphene electrons successfully describes our SPD.

\section{Methods}
\textbf{Fabrication}. Fabrication of our graphene calorimetric SPD begins with a high-resistivity silicon chip sputtered with 200 nm of niobium (Nb). Using photolithography and plasma etching, DC electrodes and a gate line are patterned from the Nb film. At the center of the pre-patterned Nb chip, a 200 $\mu$m-by-200 $\mu$m area of bare Si remains exposed for the placement of the graphene heterostructure. The hBN/graphene/hBN/graphite heterostructures are prepared and placed using standard exfoliation and stacking techniques \cite{Wang.2013}.

We use electron-beam lithography and plasma etching to define the heterostructure. The bottom graphite flake in the heterostructure serves as a gate to control the carrier density in the graphene, separated by the bottom hBN layer. This graphite layer also screens the graphene from charge inhomogeneities that may exist at the surface of the silicon. The graphite is connected using a MoRe electrode (75 nm thick), whose connection is severed from the graphene by plasma etching the top hBN and graphene.

In order to prevent short-circuiting between the graphene and graphite on either side of the heterostructure during the sputtering of the Josephson junction electrodes, both sides are insulated with a 120-nm poly(methyl methacrylate) (PMMA) layer. This layer is overdosed forming a cross-linked insulator. Afterward, the Josephson junction electrodes, made of MoRe (195 nm), are patterned by electron-beam lithography. For the primary device studied in this text, Device B, electrodes are sputtered onto a two-dimensional graphene sheet exposed by etching only the top hBN layer. For Device A, the electrodes are sputtered onto a one-dimensional graphene edge exposed by etching both the top hBN and graphene layers. Both Device A and Device B have a junction channel length of 600 nm and a width of 1.7 $\mu$m.

Lastly, electron-beam lithography is utilized to define the galvanic connection between the MoRe to the pre-patterned Nb film. To eliminate any oxide layer on the Nb and ensure superconducting contact, we employ in-situ argon ion milling before deposition. Without breaking vacuum, an adhesion layer of Ti (5 nm) is evaporated, followed by sputtering of MoRe (250 nm) on the freshly exposed surfaces. The process concludes with a lift-off in acetone to remove the excess metal. 

\textbf{Laser reflectometry}. We use laser reflectometry measurements (Fig. \ref{fig:ReflectometrySchematic}) to accurately position the beam spot onto the graphene calorimetric SPD. Described in the main text, we use a single-mode optical fiber to bring 1550-nm photons to our samples through a long-pass filter. The fiber system successfully suppresses the stray ambient light from the laboratory space down to our SPD to merely 3 photons per minute. For reflectometry, the 0.6 (0.3) numerical aperture focusing lens for device B (A) is chosen to balance between the size of the beam spot and the collection efficiency of the light reflected from the samples back to the optical fiber. After reflecting off the device, the light is routed via a directional coupler to a PbTe photodetector. We modulate the incident light intensity by applying a sinusoidal voltage bias to the laser diode. We then measure $V_{\text{refl}}$ by a lock-in amplifier at different sample positions to produce the image in Fig.~1C. The directional coupler enables continous monitoring of the incident laser power using a power meter.

Switching from laser reflectometry to single-photon measurements does not require any addition or removal of components in the optical path. We simply turn off the sinusoidal voltage bias to the diode, apply a small DC voltage bias to set the laser power output at 1 $\mu$W, and tune an in-line variable attenuator.

\textbf{$\Gamma_{\text{meas}}$ through sweeping and counting techniques}. We measure $\Gamma_{\text{meas}}$ through two different, but equivalent, measurement protocols \cite{Walsh.2020}. The first is to collect the GJJ switching statistics and extract $\Gamma_{\text{meas}}$ through the Fulton-Dunkelberger method \cite{Fulton.1974}. In this protocol, we ramp $I_b$ from -4 $\mu A$ to +4 $\mu A$ and record the junction voltage, repeated over $\sim$10$^4$ sweeps. For each sweep, we record $I_s$ at which the junction switches from superconducting to resistive. Collecting the statistics of $I_s$, we can extract a switching rate at each $I_b$ \cite{Fulton.1974}.

Complimentary to this approach is the counting method. Here, we set a constant $I_b$ below $\langle I_s \rangle$ while monitoring the voltage across the junction. When a switching event occurs, a voltage 'click' is recorded and a self-resetting circuit will bring the junction back to the superconducting state. We confirm that these two measurement techniques yield the same results \cite{Walsh.2020}.

\section{Acknowledgements}
We thank valuable discussions with J.Balgley and B.-I.Wu, and technical support from B.Hassick. Experimental setup, measurements and data analysis performed by B.H. were supported by an appointment to the Intelligence Community Postdoctoral Research Fellowship Program at the Massachusetts Institute of Technology, administered by Oak Ridge Institute for Science and Education through an interagency agreement between the U.S. Department of Energy and the Office of the Director of National Intelligence, and by E.G.A. who was supported from the Army Research Office MURI (Ab-Initio Solid-State Quantum Materials) Grant no. W911NF-18-1-043. Sample optimization, characterization, and fabrication by W.J. and G.-H.L. were supported by National Research Foundation of Korea (NRF) funded by the Korean Government (2020M3H3A1100839, RS-2024-00393599, RS-2024-00442710, RS-2024-00444725), ITRC program (IITP-2022-RS-2022-00164799) funded by the Ministry of Science and ICT, Samsung Science and Technology Foundation (SSTF-BA2101-06, SSTF-BA2401-03) and Samsung Electronics Co., Ltd (IO201207-07801-01). For hBN, K.W. and T.T. acknowledge support from the JSPS KAKENHI (Grant Numbers 21H05233 and 23H02052) and World Premier International Research Center Initiative (WPI), MEXT, Japan.



\section{Supplementary Information}
\renewcommand{\thetable}{S\arabic{table}} 
\renewcommand\thefigure{S\arabic{figure}} 
\begin{table}[t]
\centering
\begin{tabular}{ l c c } 
\hline
Device &~~~~~~A~~~~~~&~~~~~~B~~~~~~\\ 
\hline
JJ width ($\mu$m) & 1.7 & 1.7\\ 
JJ channel length (nm) & 600 & 600 \\
Graphene layer & 1 & 1 \\
MoRe thickness (nm)  &  195 & 195\\
Contact Type  &  1D & 2D\\
Graphene Target Width ($\mu$m)  &  4.8 & 4\\
Graphene Target Length ($\mu$m) &  10.6 & 25.8\\
 $\langle I_s\rangle$ ($\mu$A)& 3.3 & 3.3 \\
$V_{\text{CNP}}$ (V) & 0 & -0.15 \\
Bottom hBN thickness (nm) & 56 & 36 \\
Top hBN thickness (nm) & 28 & 51 \\
$n_{e}/V_{\text{gate}}$ (10$^{12}$ cm$^{-2}/V$) & 0.22 & 0.34 \\
Electronic mobility (cm$^{2}$/V s) & 17000 & 9100\\
Mean free path (nm) & 313 & 444 \\
$I_c$ ($\mu$A) & 3.6 & 3.38 \\
$R_n$ ($\Omega$) & 70 & 48\\
$I_cR_n$ ($\mu e$V) & 252 & 161\\
Thouless energy (m$e$V) & 1.34 & 0.91\\
JJ coupling energy (m$e$V) & 7.39 & 6.9\\
$\omega_{P0}/2\pi$ (GHz)& 199 & 131 \\
$C_{\text{JJ}}$ (fF)& 7 & 15\\
$Q(I_b = 0)$ & 0.61 & 0.59\\
$\Delta_s$ of MoRe (m$e$V) & \multicolumn{2}{c}{1.3}\\
\hline
\end{tabular}
\caption{\textbf{List of parameters for measured devices.}
$V_{\text{CNP}}$ is the gate voltage of the charge neutrality point for the monolayer graphene.}
\label{tab:Devices}
\end{table}

\subsection{Characterization of graphene-based Josephson junctions}
\begin{figure}
\includegraphics[width=\columnwidth]{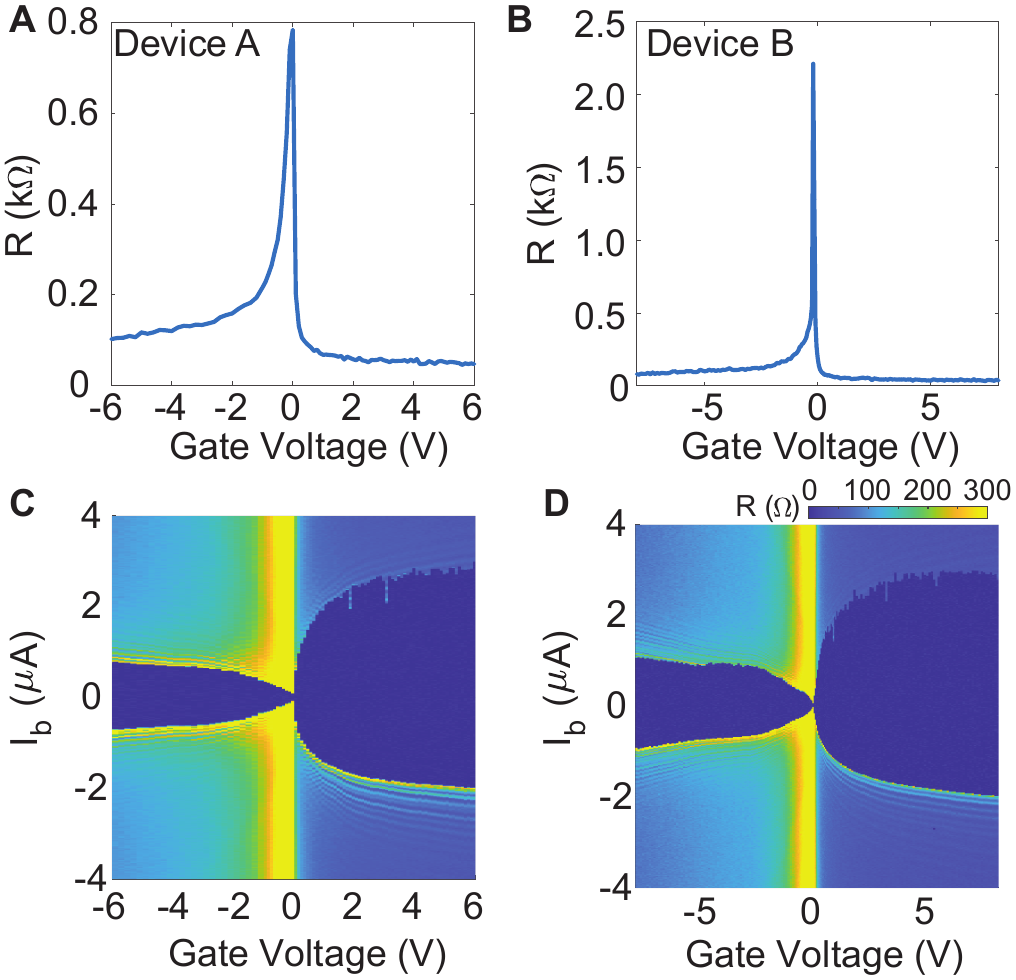}
\caption{\textbf{Characterizations of the GJJs. (A)} Device A gate dependence. \textbf{(B)} Device B gate dependence. \textbf{(C)} Device A bias-gate map. \textbf{(D)} Device B bias-gate map.}
\label{fig:BasicCharacterization}
\end{figure}

Fig. \ref{fig:BasicCharacterization} shows the $I-V$ characteristics of the GJJs. Fig. \ref{fig:BasicCharacterization} A and B plots $R_n(V_{\text{gate}})$ with $I_b$ of 4 $\mu$A $> \langle I_s\rangle$. Both devices have a sharp, singular charge neutrality point at 0 V and -0.15 V for devices A and B, respectively. This indicates that the charge density is homogeneous in the junction region.

Fig. \ref{fig:BasicCharacterization}C and D show the full scan of $R_n(I_b, V_{\rm{gate}})$. The robust supercurrent that flows on either side of the charge neutrality point\cite{Heersche.2007,Du.2008,Shalom.2015,Lee.2011} is consistent with previous reports in using MoRe to fabricate GJJs \cite{Borzenets.2016nj,Calado.2015}. This is attributed to the minimal doping caused by the MoRe on the graphene. Collectively, these measurements indicate high-quality GJJs that can support a wide range of supercurrents for the calorimetric SPD. Table \ref{tab:Devices} summarizes the characteristics of the two measured devices.

\subsection{Calculating the reflectance and absorption coefficient of hBN-encapsulated graphene}\label{secAbsorption}
The optical properties of a material depends on its dielectric environment. Similar to superconducting nanowire detectors and transition edge sensors \cite{Holzman.2019,Gaggero.2010,Reddy.2020,Korzh.2020,Irwin.Hilton.2005,Patel.Thomas.2024}, we can improve the efficiency of our calorimetric SPD in the future by optimizing the absorption coefficient of incident photons of the graphene \cite{Gan.2012,Furchi.Mueller.2012,Vasic.2014,Efetov.2018}. For this report, we calculate the reflectivity of the entire graphene heterostructure, $\mathcal{R_{\text{hs}}}$, and the graphene absorption coefficient, $\alpha_{\text{gr}}$, using the wave-transfer matrix method \cite{Deng.2015,SensaleRodriguez.2012,Mehew.2024,Gaggero.2010}. We find excellent agreement between the measured and calculated ratio of $\mathcal{R_{\text{hs}}}$ on silicon to the graphene heterostructures (Fig.~2A).

\label{sec:waveMatrix}
\begin{figure}
\begin{center}
\includegraphics[width=0.5\columnwidth]{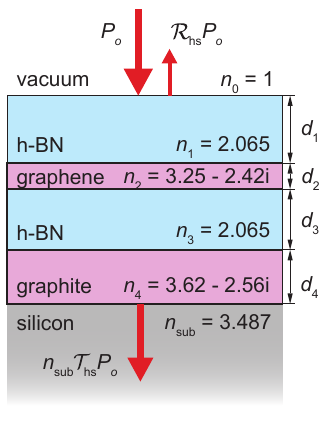}
\end{center}
\caption{\textbf{Schematic of the graphene heterostructure in our calorimetric SPD.} Side view of the heterostructure used to calculate the 1550 nm absorption and reflection values of the graphene calorimetric SPDs. From top to bottom, the heterostructure consists of alternating h-BN and graphene/graphite layers terminating with a semi-infinite layer of Si. In our calculations, $P_0$ is the optical power incident from vacuum onto the heterostructure. The power that is transmitted through the heterostructure is $n_{\text{sub}}\mathcal{T_{\text{hs}}}P_o$ whereas the power that is reflected from the heterostructure is $\mathcal{R_{\text{hs}}}P_o$. The red arrows denote the Poynting vector associated with incident, reflected, and transmitted light.}
\label{fig:ThinFilmStack}
\end{figure}

In the wave-transfer matrix method, $M_i$ and $M_{i,j}$ are the matrices describing, respectively, the phase accumulation of light when traverses through the $i$-th optical layer, and the transmission and reflection at the interface when the light traversing from the $j$-th to the $i$-th optical layer \cite{Saleh.2019}:

\ba
    M_i &=& \left( \begin{array}{cc}
    \mathrm{exp}(-2{\pi}jn_id_i / \lambda) & 0 \\ 0 & \mathrm{exp}(2{\pi}jn_id_i / \lambda)
    \end{array}
    \right)\\
    M_{i,j} &=& \frac{1}{2n_j} \left( \begin{array}{cc}
    n_j+n_i & n_j-n_i \\ n_j-n_i & n_j+n_i
    \end{array}
    \right)
\ea with $\lambda \simeq$ 1550 nm being the wavelength of light, $n_i$ the index of refraction, and $d_i$ thicknesses of the $i$-th layer.

Fig.~\ref{fig:ThinFilmStack} depicts the graphene heterostructure in our calorimetric SPD with the index assignment for each layer. The overall matrix, $M$, of the hBN/graphene/hBN/graphite heterostructure on a semi-infinite substrate denoted with the subscript ``sub'' is given by:
\begin{equation}
    M = M_{\text{sub},4}\prod_{i=1}^{4}M_iM_{i,i-1} = \left( \begin{array}{cc}
    m_{11} & m_{12} \\ m_{21} & m_{22}
    \end{array}
    \right)
\end{equation}

For light of an electric field amplitude in the $i$-th layer, $E^\beta_i$, with $\beta = +(-)$ denoting the light propagation from vacuum to substrate (substrate to vacuum), we have: 
\ba
\left( \begin{array}{c} E^+_{\text{sub}} \\ 0 \end{array} \right) &=& M \left( \begin{array}{c} E^+_0 \\ E^-_0 \end{array} \right).
\label{eqn:waveMatrix_fullstack}\\
    E^-_0 &=& -\frac{m_{21}}{m_{22}}E^+_0
\label{eqn:E0-}\\
    E^+_{\text{sub}} &=& \left( m_{11}-\frac{m_{12}m_{21}}{m_{22}} \right) E^+_0.
\label{eqn:Esub+}
\ea
\begin{table}
\centering
\begin{tabular}{ l c c c c c c} 
\hline
Device &$d_1$ (nm)&$d_2$ (nm)&$d_3$ (nm)&$d_4$ (nm)& $\alpha_{\text{gr}}$ ($\%$)&$\alpha_0$ ($\%$)\\ 
\hline
A & 28 & 0.33 & 56 & 2& 0.62&2.3\\ 
B & 51 & 0.33 & 36 & 1.33 & 0.61&2.3\\
\hline
\end{tabular}
\caption{\textbf{Thickness parameters and calculated absorbance values of GJJ devices. }Absorbance calculations are performed assuming 1550 nm photons that are normally incident on the GJJ devices.
}
\label{tab:ThinFilmThicknesses}
\end{table}
\begin{table}
\centering
\begin{tabular}{ l c } 
\hline
Material &~~~$\mathcal{R}$~~~\\ 
\hline
Si & 0.31\\ 
Device A & 0.21\\
Device B & 0.21\\
\hline
\end{tabular}
\caption{\textbf{Calculated $\mathcal{R}$ values of 1550 nm photons at normal incidence from vacuum.}
$\mathcal{R}$ of the relevant materials on the graphene calorimetric SPD chip at a photon wavelength of 1550 nm. We calculate the $\mathcal{R}$ of Si from the Fresnel equations, $\mathcal{R}=|(1-n)/(1+n)|^2$, where $n$ is the refractive index of Si. We calculate $\mathcal{R}$ of the graphene calorimetric SPDs using Eqn. \ref{eqn:Rstack} from the wave-transfer matrix method. We experimentally determine $\mathcal{R}$ of MoRe from reflectometry measurements to be $\sim0.85$.}
\label{tab:Reflectances}
\end{table}

Now we calculate the reflectivity of the heterostructure, $\mathcal{R_{\text{hs}}}$. The flow of optical power is determined by Poynting vectors, $S^{\beta}_i$: 
\begin{equation}
S^{\beta}_i = \text{Re}(E^\beta_i H^{*\beta}_i)
\label{eqn:poynting}
\end{equation}
where $H^\beta_i$ is the magnetic field amplitude of light traveling in the $i$-th optical layer in the $\beta$ direction. The electric and magnetic fields are related by $H^\beta_i = n_i E^\beta_i / Z_0$, with $Z_0 = \sqrt{\mu_0 / \epsilon_0}$ being the free-space impedance, $\epsilon_0$ the vacuum permittivity, and $\mu_0$ the vacuum permeability. Hence, $S^+_0$, $S^-_0$ and $S^+_{\text{sub}}$ are given by:
\ba
S^+_0 &=& \frac{|E^+_0|^2}{Z_0}\label{eqn:S+0}\\
S^-_0 &=& \left|-\frac{m_{21}}{m_{22}} \right|^2 S^+_0\label{eqn:S-0}\\
S^+_{\text{sub}} &=& n_{\text{sub}} \left|  m_{11}-\frac{m_{12}m_{21}}{m_{22}} \right|^2 S^+_0
\label{eqn:S+sub}
\ea
We can calculate $\mathcal{R_{\text{hs}}}$ using:
\ba
    \mathcal{R_{\text{hs}}} &=& \frac{S^-_0}{S^+_0} = \left|-\frac{m_{21}}{m_{22}} \right|^2
\label{eqn:Rstack}
\ea and the absorption coefficient, $\alpha_l$, of the $l$-th layer using \cite{Deng.2015,Mehew.2024}:
\begin{equation}
\alpha_l = \frac{(S^+_0 - S^-_0) - (S^+_l - S^-_l)}{S^+_0}
\end{equation}
Using the partial wave-transfer matrix, $X$:
\ba
\left( \begin{array}{c} E^+_{\text{sub}} \\ 0 \end{array} \right) &=& X \left( \begin{array}{c} E^+_l \\ E^-_l \end{array} \right)
\label{eqn:waveMatrix_partialstack}\ea where \ba
    X &=& \begin{cases} 
        M_{\text{sub},4}M_4 \left( \prod_{i=l}^{3}M_{i+1,i}M_i \right) & 1 \leq l \leq 3 \\
        M_{\text{sub},4}M_4 & l = 4
    \end{cases} \\
     &=& \left( \begin{array}{cc}
    x_{11} & x_{12} \\ x_{21} & x_{22}
    \end{array} \right)
\label{eqn:partialWaveMatrix}
\ea we have:
\ba
S^+_{l} &=& n_{l} \left| \frac{m_{11}-m_{12}m_{21}/m_{22}}{ x_{11}-x_{12}x_{21}/x_{22}} \right|^2 S^+_0
\label{eqn:S+l}\\
S^-_{l} &=& n_{l} \left| \frac{m_{11}-m_{12}m_{21}/m_{22}}{ x_{12}-x_{11}x_{22}/x_{21}} \right|^2 S^+_0
\label{eqn:S-l}
\ea
\begin{widetext}
\begin{equation}
\alpha_l = 1 - \left|\frac{m_{21}}{m_{22}} \right|^2 - n_l \left( \left| \frac{m_{11}-m_{12}m_{21}/m_{22}}{ x_{11}-x_{12}x_{21}/x_{22}} \right|^2 - \left| \frac{m_{11}-m_{12}m_{21}/m_{22}}{ x_{12}-x_{11}x_{22}/x_{21}} \right|^2 \right)
\end{equation}
\end{widetext}
Using the measured thickness listed in Table \ref{tab:ThinFilmThicknesses}, we find that $\alpha_{\text{gr}} \approx 0.6 \%$, lower than the free-standing value of $2.3\%$. Although our experimental setup would not allow us to directly measure $\alpha_{\text{gr}}$, we find that the wave-transfer matrix method provides an excellent agreement on the calculated ratio of the reflectance values of the graphene heterostructure to silicon (Table \ref{tab:Reflectances}) with the measured value (gray dashed lines, Fig. 2A).
\begin{figure*}
\includegraphics[width=1.4\columnwidth]{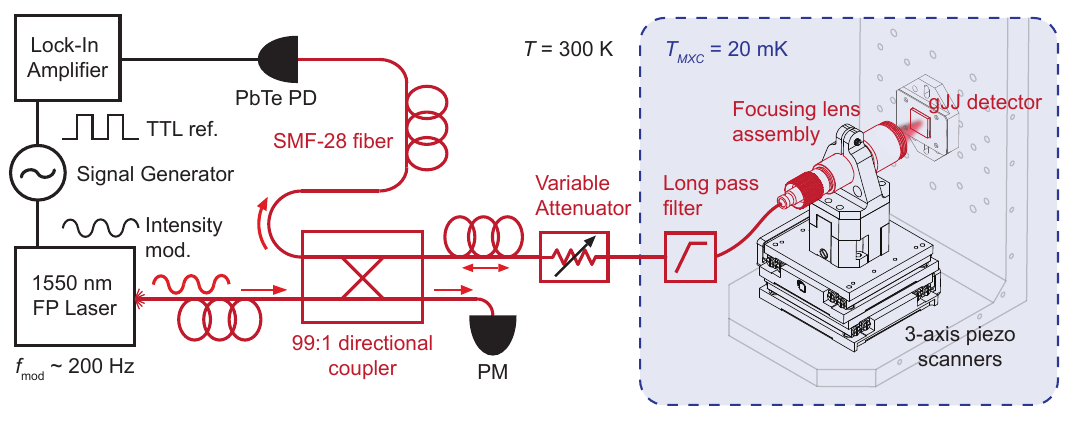}
\caption{\textbf{Reflectometry schematic.} Schematic of the reflectometry setup used to locate and position our laser spot onto the graphene. Laser light from an intensity-modulated 1550 nm Fabry-Perot (FP) laser is routed through an SMF-28 single-mode fiber onto the graphene calorimetric SPD. This light is then reflected off the detector and routed to a PbTe photodetector (PD) that transduces an AC-modulated electrical signal proportional to the reflected light intensity hitting the PbTe PD. The AC-modulated electrical signal is then demodulated and amplified using a lock-in amplifier. The RMS power of our laser light as measured by an optical power meter (PM) is $\sim$200 $\mu$W. Upon the 20 dB attenuation passing through the directional coupler and roughly 3 dB of loss from insertion loss between fiber components and the focusing lens assembly, we estimate that 1 $\mu$W RMS of laser power is incident on the graphene calorimetric SPD during the reflectometry measurements.}
\label{fig:ReflectometrySchematic}
\end{figure*}
\begin{figure*}
\includegraphics[width=1.4\columnwidth]{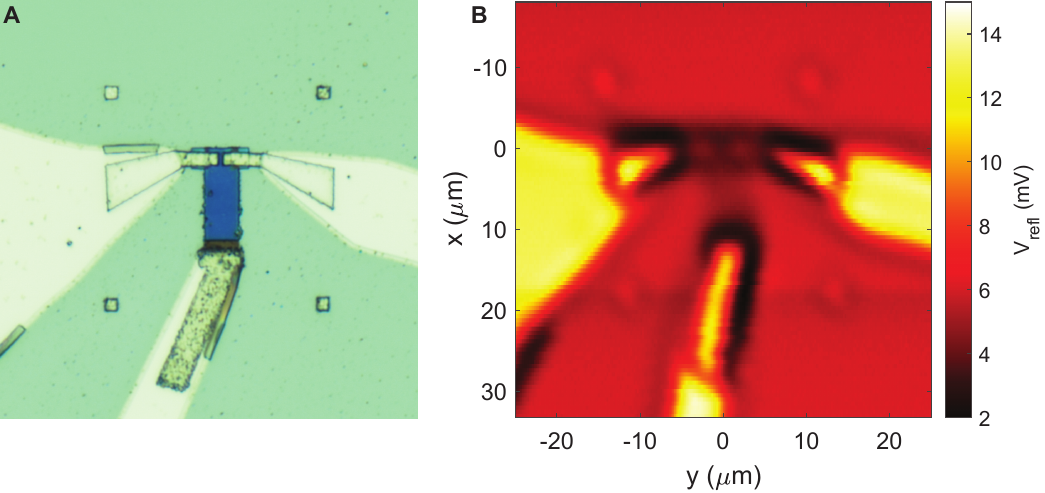}
\caption{\textbf{Reflectometry measurement used to find the graphene target. (A)} Optical image of Device A. \textbf{(B)} 2D reflectometry scan of Device A taken at 140 mK.}
\label{fig:FindingGraphene}
\end{figure*}

\subsection{Scanning laser reflectometry and estimation of the beam spot size}
We use the laser reflectometry setup (Fig.~\ref{fig:ReflectometrySchematic}) described in the Methods section to locate our beam spot with respect to our device. Fig. \ref{fig:FindingGraphene} compares an optical image of Device A with our scanned reflectometry image taken at 140 mK. The highly reflective superconducting MoRe electrodes and gate leads are clearly discernible as bright yellow regions in Fig. \ref{fig:FindingGraphene}B. Below the junction, the graphene heterostructure appears as the darker red region. These features enable us to precisely place the beam spot on our graphene calorimetric SPD.

Experimentally, we determine the beam spot size, $2w_o$, from $V_{\text{refl}}$ in Fig. \ref{fig:FindingGraphene}B and Fig. \ref{fig:DeviceAFig2}A. $V_{\text{refl}}(x,y)$ images the graphene heterostructure with a point spread function given by the Gaussian beam profile of the beam spot, $I(x,y)$, through a single-mode optical fiber such that:
\begin{equation}
V_{\text{refl}}(x,y) \propto\iint{\mathcal{R}(x',y')I(x-x',y-y')\mathrm{d}x'\mathrm{d}y'}
\label{eqn:reflectanceConvolution}
\end{equation} where $\mathcal{R}$ is the reflectance, and 
\begin{equation}
I(x,y) = I_0\mathrm{exp} \left( \frac{-8(x^2+y^2)}{(2w_o)^2} \right)
\end{equation} 
with $I_0 = 8P_{\mathrm{laser}}/{\pi}(2w_o)^2$ being the intensity at the center of the beam spot, and $P_{\mathrm{laser}}$ being the laser power through the optical fiber. Using the values of $\mathcal{R}$ for hBN-encapsulated graphene and silicon calculated in Table \ref{tab:Reflectances}, we fit the $V_{\text{refl}}$ data in Fig. \ref{fig:Fig2}A and Fig. \ref{fig:DeviceAFig2}A with the convolution integral (Eqn. \ref{eqn:reflectanceConvolution}) for the beam spot size. The best fitted values of $2w_o$ are given in Table \ref{tab:beamSpotSizes}.

This beam spot size, however, is larger than the diffraction limit, $2w^{\rm{dl}}_o$, given by:
\begin{equation}
2w^{\rm{dl}}_o = \frac{\lambda}{2\mathrm{NA}}
\label{eqn:diffractionLimit}
\end{equation} where NA is the numerical aperture of the focusing lens. We can reconcile the discrepancy by considering the underfilling of light through the focusing lens\cite{Marom.1979,Urey.2004}. We calculate the enlarged beam spot size\cite{Urey.2004}, $2w_o'$:
\begin{equation}
2w_o' = K \lambda f_\#
\end{equation}
where $f_\#$ is the f-number, defined as the ratio between the focal length and diameter, $\phi_{\text{lens}}$, of the focusing lens. The beam spot constant, $K$, is given by\cite{Urey.2004}:
\begin{equation}
K = 1.654-\frac{0.105}{\tilde{\phi}}+\frac{0.28}{\tilde{\phi}^2}
\end{equation}
where $\tilde{\phi} \equiv \phi_{\text{coll}}/\phi_{\text{lens}}$ is the truncation ratio, with $\phi_{\text{coll}}$ being the diameter of collimated light entering the focusing lens. For both GJJ calorimetric SPD devices, $\phi_{\text{coll}} = 3.6$ mm. The agreement between the extracted beam spot size and our calculation of $2w_o'$ for both devices suggests that the underfilling of the lens causes the beam spot size in our experiment larger than the diffraction limit.

\begin{table}
\centering
\begin{tabular}{l c c}
\hline
Device &~~~~~~A~~~~~~&~~~~~~B~~~~~~\\
\hline
NA & 0.3 & 0.6 \\
$f_\#$& 1.7 & 0.8 \\
$\phi_{\text{lens}}$ (mm)& 5.5 & 6.5 \\
$\phi_{\text{coll}}$ (mm) & 3.6 & 3.6 \\
$\tilde{\phi}$ & 0.65 & 0.55 \\
$K$ & 2.2 & 2.4 \\
diffraction-limited $2w^{\text{dl}}_o$ ($\mu$m)& 2.6& 1.3\\
finite-size $2w'_o$ ($\mu$m) & 5.7 & 3.1 \\
fitted $2w_o$ ($\mu$m) & 7& 4\\
\hline
\end{tabular}
\caption{\textbf{Optical system parameters and beam spot sizes.} Collimating and focusing lens parameters used in the calculation of the enlarged beam spot size, $2w^\prime_o$, for both GJJ calorimetric SPD devices.}
\label{tab:beamSpotSizes}
\end{table}

\subsection{Estimating the quantum efficiency}
Experimentally, we infer $\eta$ in Fig. 1D, 2C and 2F by $\eta = \Gamma_{\text{meas}}/\dot{\mathcal{N}}_{\text{abs}}$, where $\dot{\mathcal{N}}_{\text{abs}}$ is the photon rate absorbed by the graphene. $\dot{\mathcal{N}}_{\text{abs}}$ is the product of the incident photon rate $\dot{\mathcal{N}}$ and $\alpha_{\text{gr}}$, i.e. $\dot{\mathcal{N}}_{\text{abs}} = \alpha_{\text{gr}} \dot{\mathcal{N}}$. The incident photon rate is determined by both the spatial overlap of the laser spot with the graphene, and the laser power that is applied on the device. As a function of position, $\dot{\mathcal{N}}(x,y)$ is calculated by performing a convolution between the laser intensity, $I(x,y)$, with the graphene binary profile, $\mathcal{G}(x,y)$, normalized to the energy of a single photon, $h\nu$: 
\begin{equation}
\dot{\mathcal{N}}(x,y) = \frac{1}{h\nu}\iint{\mathcal{G}(x',y')I(x-x',y-y')\mathrm{d}x'\mathrm{d}y'}\label{eqn:ndot}
\end{equation} where $\mathcal{G}(x,y) = 1$ on the graphene heterostructure, and 0, otherwise. When the beam spot is centered on the middle of the graphene heterostructure, with $P_{\text{laser}} = 1$ fW, $\dot{\mathcal{N}} \simeq$~7.3 kHz (Device B) and $\dot{\mathcal{N}}_{\text{abs}}\simeq$~45 Hz based on the value of $\alpha_{\text{gr}}$ in Table \ref{tab:ThinFilmThicknesses}. In the $I_b$ ranges where $\Gamma_{\text{meas}}$ plateaus at about 35 $\pm$ 4 Hz in Fig. 1D, $\eta \simeq$ 0.78 $\pm$ 0.08.

\subsection{Estimation of the Impact of a Plasmon Mode}
Previous work on GJJ SPDs relied on the presence of a plasmon mode at the NbN-graphene interface to enhance the photon absorption efficiency \cite{Walsh.2021}. While the detection mechanism in the previous experiment is through Cooper pair breaking rather than the calorimetric effect in this report, here we consider how the presence of a plasmon mode, if it exists, would impact $\Gamma_{\text{meas}}$. To begin, we calculate the number of incident photons per unit time per unit area per unit incident laser power through the focusing lens, $\mathcal{J}_{\text{photon}}$, given by:
\begin{equation}
\mathcal{J}_{\text{photon}} = \dot{\mathcal{N}}/\pi w_o^2
\end{equation} with $\dot{\mathcal{N}}$ being the photon rate in the Gaussian beam and $\pi w_o^2$ the beam spot area. We find $\mathcal{J}_{\text{photon}} = $ 581 photons per second per $\mu$m$^2$ per 1 fW of laser power.

If a similar plasmonic mode exists, based on Ref. \cite{Walsh.2020}, we can estimate an effective single-photon absorption area, $\mathcal{A}_{\rm{eff}}$, and averaged photon absorption coefficient, $\langle\alpha\rangle_{\rm{plasmon}}$, to calculate the expected absorbed photon rate due to a plasmon, $\dot{\mathcal{N}}^{\rm{(plasmon)}}_{\text{abs}}$:
\begin{equation}
    \dot{\mathcal{N}}^{\rm{(plasmon)}}_{\text{abs}} = \mathcal{J}_{\text{photon}} \mathcal{A}_{\rm{eff}} \langle\alpha\rangle_{\rm{plasmon}} P_{\text{laser}}
\end{equation}
We find that $\dot{\mathcal{N}}^{\rm{(plasmon)}}_{\text{abs}}$ would be 371 Hz. This value is $\sim$8 times more efficient than direct absorption by the graphene layer (Table \ref{tab:Plasmon}), which would result in a sizable enhancement in $\Gamma_{\text{meas}}$ when the beam spot is centered on the GJJ. Instead, we observe a reduced $\Gamma_{\text{meas}}$ in Fig.~2, suggesting that the plasmon mode is either substantially less effective than in Ref. \cite{Walsh.2020} or not present in the graphene-MoRe interface. We attribute the difference to the graphene-superconductor interface --- the superconducting electrodes are made of MoRe in this report rather than NbN in the previous one. Regardless, the data suggests that plasmonics play little to no role in our experiment. 

\begin{table}[t]
\centering
\begin{tabular}{ l c c c c} 
\hline
Reference &~~~$\mathcal{A}_{\text{eff}}$ ($\mu$m$^2$)~~~&~~~$\langle \alpha\rangle$~~~&~~~$\dot{\mathcal{N}}_{\text{abs}}$ (Hz)~~~\\ 
\hline
Plasmon & 2 × 2.8 × 0.19 & 0.6 & 371\\
Graphene & 4$\pi$ & 6.1 × $10^{-3}$ & 45\\
\hline
\end{tabular}
\caption{\textbf{Comparison of single-photon absorption rate between the plasmon mode and graphene.}
Relevant parameters for the calculation of $\dot{\mathcal{N}}_{\text{abs}}$. In the calculations of $\dot{\mathcal{N}}_{\text{abs}}$ from both the GJJ and the plasmon mode, $\mathcal{J}_{\text{photon}} = 581$ photons per second per $\mu$m$^2$ per 1 fW, and $P_{\text{laser}} =$ 1 fW.}
\label{tab:Plasmon}
\end{table}

\subsection{Single-Photon Detection in a Second Device}

\begin{figure*}
\includegraphics[width=1.5\columnwidth]{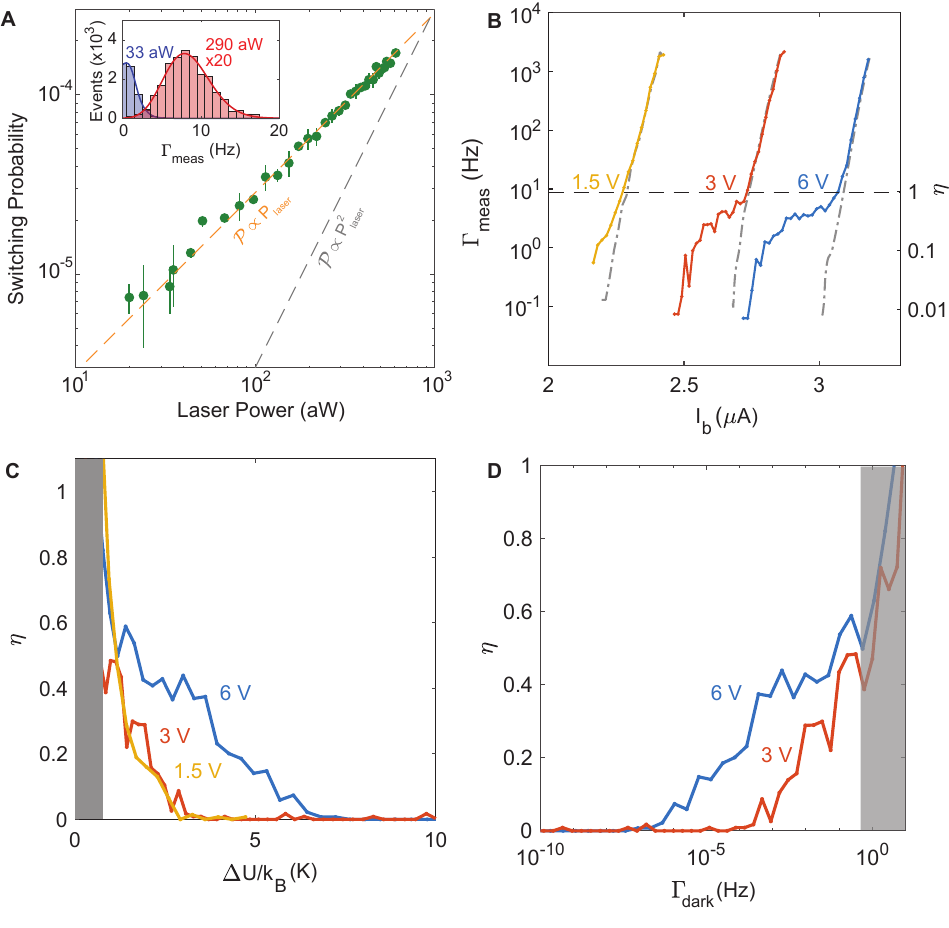}
\caption{\textbf{Single-photon signal in Device A. (A)} Switching probability as a function of applied laser power. Linearity indicates that the switching events are caused by a single photon. Inset: The switching events adhere to Poissonian statistics indicating that the device is shot-noise limited. \textbf{(B)} $\Gamma_{\text{meas}}(V_{\rm{gate}})$ for Device A with $P_{\rm{laser}} = 200$~aW. \textbf{(C)} $\eta$ vs. $\Delta U/k_B$ shows that at the highest $V_{\text{gate}} = 6$ V, a single photon can induce the escape of the GJJ phase particle from a $\Delta U/k_B$ of $\sim$ 2 K with $\eta = 0.5$. Gray box indicates the region where self-switching of the junction is dominant. \textbf{(D)} Tradespace between $\eta$ vs $\Gamma_{\text{dark}}$ shows Device A can detect a single photon with $\eta \sim 0.5$ for $\Gamma_{\text{dark}}$ at 1 Hz. Gray box indicates the region where self-switching of the junction is dominant.}
\label{fig:DeviceAFig1}
\end{figure*}

\textit{(Single-photon sensitivity.)} In addition to the device which was studied in the main text, we have also confirmed single-photon detection in a second device (Device A). Parameters for both devices are listed in Table \ref{tab:Devices}. We confirm the single-photon detection in the same manner as the main text by placing the beam spot 5 $\mu$m from the junction and applying a fixed $I_b$ at $\sim$78\% of $I_c$ while measuring $\Gamma_{\rm{meas}}$. As with Device B, we find that the switching events adhere to Poissonian statistics (Fig. \ref{fig:DeviceAFig1}A inset), indicating the detector is shot-noise limited. We plot the switching probability against $P_{\rm{laser}}$ and find a linear trend (Fig. \ref{fig:DeviceAFig1}A), indicating that the switching events are due to the detection of a single photon.

\textit{(Gate dependence.)} In order to compare the two devices, we explore the performance of Device A on electron density. As $V_{\rm{gate}}$ increases from the charge neutrality point ($\simeq$0.0 V), $\Gamma_{\rm{meas}}$ becomes considerably larger than $\Gamma_{\rm{dark}}$ (Fig. \ref{fig:DeviceAFig1}B), similar to the data from Device B shown in Fig. 3. The nonlinear $\Gamma_{\rm{meas}}$ in the log-linear plot in Fig.~\ref{fig:DeviceAFig1}B indicates that $\Gamma_{\rm{meas}}$ is not proportional to $\sim\exp{-\Delta U}$, i.e. in activation theory. Comparison of the experimental data and theory \cite{Lee.2020,Walsh.2021} shows that this switching behavior of a Josephson junction is due to discrete triggering event, i.e. single photons, rather than the induction from a thermal bath in equilibrium. This is an important distinction of our graphene calorimetric SPD \cite{Walsh.2017} from graphene bolometers \cite{Yan.Fuhrer.2012,Fong.2012,Vora.Du.2012,Du.2014,Fatimy.2016,Efetov.2018,Lee.2020,Kokkoniemi.2020,Cai.2014,Blaikie.2019,Skoblin.2018,Yuan.2020,Han.2013,Hruby.Neugebauer.2024,Sassi.2017}.

Up to the highest attainable $V_{\rm{gate}}$ in Device A (the bottom hBN layer is 20 nm thicker than Device B), $\Gamma_{\text{meas}}$ does not fully saturate. This is despite the $n_e$ explored for Device A falling within the range of $n_e$ that provided robust bias saturation in Device B. Moreover, $\eta$ for Device A is considerably less than that of Device B. The correlation between of the non-saturating $\Gamma_{\text{meas}}(I_b)$ and a lower $\eta$ is consistent with previous studies \cite{Walsh.2021,Battista.Efetov.2024}.

Similar to Fig.~3D and E for Device B, we study the calorimetric effect of and $\eta$-vs.-$\Gamma_{\rm{dark}}$ tradespace of Device A. Fig.~\ref{fig:DeviceAFig1}C plots $\eta$ vs. $\Delta U/k_B$ at several $V_{\text{gate}}$. At the highest gate voltage, $V_{\text{gate}} = 6$ V, a single photon can induce phase-particle escape from a $\Delta U/k_B \sim$ 2 K with $\eta \sim 0.5$. We also plot $\eta$ vs $\Gamma_{\text{dark}}$ in Fig.~\ref{fig:DeviceAFig1}D and find that at the same gate voltage, Device A can detect a single photon with $\eta \sim 0.5$ for $\Gamma_{\text{dark}}$ at 1 Hz. These figures of merit are lower than that of Device B. 

\begin{figure*}
\includegraphics[width=1.2\columnwidth]{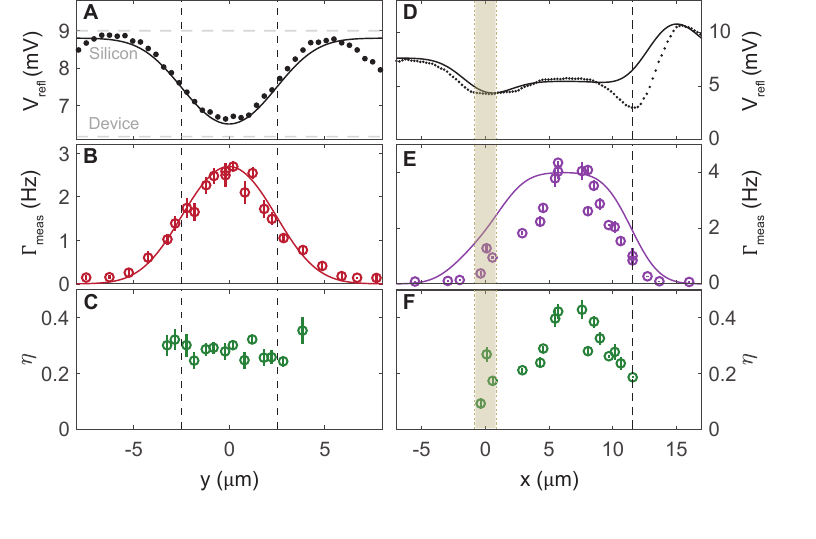}
\caption{\textbf{Scanning the photon input position across the graphene (A-C)} in the transverse ($y$-) direction. Vertical dashed lines marks the graphene location. \textbf{(D-F)} Scanning in longitudinal ($x$-) direction. Vertical purple region designates the Josephson-junction location. Vertical, gray dashed line marks the graphene location.  \textbf{(A, D)} The transverse reflectance signal (dots) and fitted step response based on the dielectric constants of graphene and silicon, interference effects, and geometric convolution assuming a Gaussian beam of spot size, 7 $\mu$m (solid line). The decrease in reflectance signal in \textbf{(A)} at the extremities of the transverse scan result from the beam spot clipping the edge of the nearby MoRe electrodes. \textbf{(B, E)} The measured switching rate (open circles) and expected switching rate based off of the step response (solid line). \textbf{(C, F)} The extracted $\eta$ (open circles) when the beam is over the device. }
\label{fig:DeviceAFig2}
\end{figure*}

\textit{(Spatial scanning.)} To further explore this difference in single-photon detection between the devices, we study the spatial dependence of Device A. As in the main text, we position the center of our beam spot $\sim$5~$\mu$m below the edge of the superconducting contact and scan the laser along the y-direction of the graphene heterostructure. The incident laser power is fixed at 200 aW. As we traverse the graphene target, the reflectance signal decreases (\ref{fig:DeviceAFig2}A) and $\Gamma_{\text{meas}}$ increases (Fig. \ref{fig:DeviceAFig2}B) as the beam spot moves towards the center of the graphene. When the spot is centered on the graphene, $\Gamma_{\text{meas}}$ reaches a maximum and plummets when the beam spot scans off the opposite edge of the graphene. We calculate $\dot{\mathcal{N}}_{\text{abs}}$ using the convolution integral (Eqn.~\ref{eqn:ndot}) of $\dot{\mathcal{N}}$ and $\alpha_{\text{gr}}$ in Table~\ref{tab:ThinFilmThicknesses} and retrieve $\eta$ as a function of position (Fig. \ref{fig:DeviceAFig2}C).

We now turn to scanning the laser across the longitudinal extent of the device. The beam is centered along the horizontal extent of the graphene target and scanned along the x-direction of \ref{fig:DeviceAFig1}. We measure the reflectance signal (Fig. \ref{fig:DeviceAFig2}D), along with $\Gamma_{\text{meas}}$ as a function position away from the GJJ (denoted by shaded yellow region in Fig. \ref{fig:DeviceAFig2}D-F). Again, we find that geometric effects dominate: $\Gamma_{\text{meas}}$ is predominantly proportional to the geometric overlap between the beam and graphene target. It is important to note that Device A is roughly half the longitudinal length of Device B. Therefore, we do not see the flat plateau in $\Gamma_{\text{meas}}$ observed in Fig. \ref{fig:Fig2}. 

Fig. \ref{fig:DeviceAFig2}F shows $\eta$ in the longitudinal direction after accounting for the overlap of the beam spot with the graphene heterostructure (Eqn.~\ref{eqn:ndot}). When the beam spot is positioned over the leads we observe a depression in $\eta$. Notably, we find that along the longitudinal extent of Device A, $\eta$ never reaches the $\eta$ observed in Device B. While the exact reason for this observation is beyond the scope of this manuscript, we note that the operation $n_e$, $\langle I_s \rangle$ and $I_r$ between Devices A and B are roughly the same. Therefore, we speculate that the device design is the primary reason for the comparatively smaller $\eta$ in Device A. Notably, Device A uses 1D contacts, while Device B uses 2D contacts. It is possible that the different fabrication methods may result in different densities of resonant scatterers\cite{Halbertal.2017} at the graphene-superconductor interface, which could increase the cooling rate of the graphene electrons. Experiments using different fabrication methods and different sizes of graphene absorber will need to optimize the SPD performance.

\subsection{Extended data on the Poisson Statistics}
\begin{figure*}
\includegraphics[width=1.2\columnwidth]{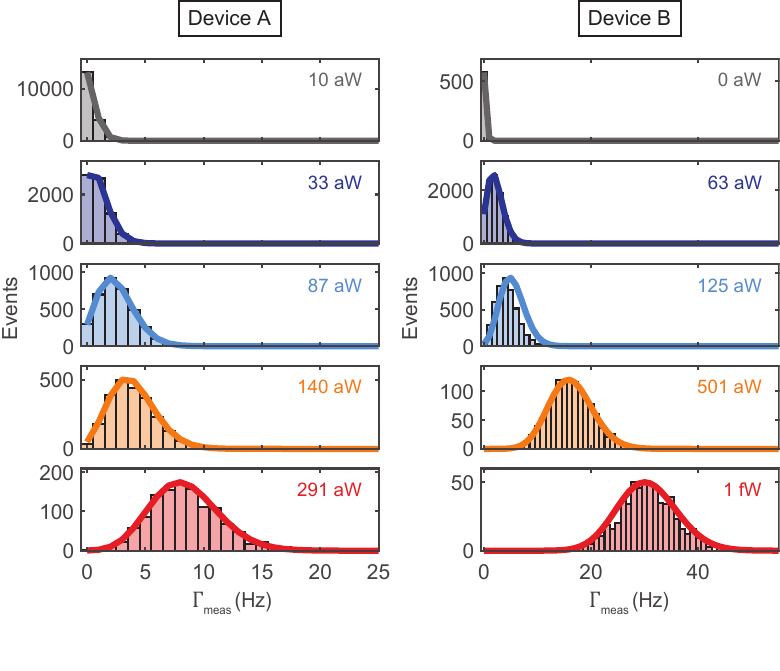}
\caption{\textbf{Extended Poissonian statistics for both devices.} The distributions of time-binned switching events with several applied laser powers adhere to Poissonian statistics, indicating that our devices are shot-noise limited. }
\label{fig:ExtendedPoisson}
\end{figure*}

As mentioned in the main text, the time-binned switching events measured at various laser powers adhere to Poissonian statistics. This indicates that our devices are shot-noise limited. In Fig. \ref{fig:ExtendedPoisson}, we show the extended distributions for several laser powers for both devices. The solid lines are the Poissonian fit. 

\subsection{Single-photon detection at 1.2 K}

\begin{figure}
\includegraphics[width=0.9\columnwidth]{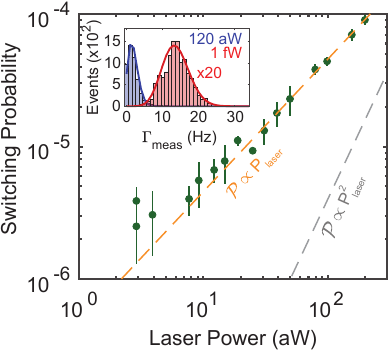}
\caption{\textbf{Single photon signal at 1.2 K.} Linearity of the switching probability as a function of laser power. \textbf{Inset} Poisson distributions show the device is still shot-noise limited at 1.2 K}
\label{fig:HighTempSPD}
\end{figure}

In order to verify SPD at elevated temperatures, we repeat the procedure described in the main text. At $T_0 =$~1.2 K, we again apply a fixed $I_b$ to the junction at roughly 90\% of $I_c$ and measure $\Gamma_{\text{meas}}$. As with low temperature, we find that $\Gamma_{\text{meas}}$ adheres to Poissonian statistics (Fig. \ref{fig:HighTempSPD} inset), indicating the detector is still shot-noise limited at 1.2 K. We plot $\Gamma_{\text{meas}}$ against $P_{\rm{laser}}$ and find a linear trend (Fig. \ref{fig:HighTempSPD}), indicating that the switching events are due to the detection of a single photon. Therefore, our device operates as an SPD up to 1.2 K.

\subsection{Macroscopic Quantum Tunneling Fits of the Junction}

To show that in the absence of illumination our device is nominally operated in the MQT regime, we fit $\Gamma_{\text{meas}}$ as a function of $I_b$ to the MQT model. Specifically, our switching rate is described as $\Gamma_{\text{meas}} = A\exp({-\Delta U}/{k_B T_{\text{eff}}})$, where, in the MQT regime \cite{Devoret.1985}:

\begin{equation}
A = A_{\text{MQT}} = 12 \omega_p \sqrt{\frac{3\Delta U}{2\pi\hbar\omega_p}}
\end{equation}
and
\begin{equation}
T_{\text{eff}} = \hbar \omega_p / [7.2 k_B (1+0.87/Q)]
\end{equation} where $\omega_p = \omega_{p0}(1-\gamma_{\text{JJ}}^2)^{\frac{1}{4}}$ is the junction plasma frequency, $\omega_{p0}= ({2eI_c}/{\hbar C_{\text{JJ}}})^\frac{1}{2}$ is the zero bias plasma frequency, $C_{\text{JJ}} = \hbar/R_n E_{\text{Th}}$ is the junction shunting capacitance \cite{Lee.2011}, with $R_n$ as the normal state resistance, and $E_{\text{Th}} = \hbar \mathcal{D}/L^2$ as the Thouless energy, where $L$ is the channel distance of GJJ, $\mathcal{D} = v_F l_{\text{mfp}}/2$ is the diffusion constant, with $l_{\text{mfp}}$ as the mean free path,  $\gamma_{\text{JJ}} = I_b/I_c$ is the normalized bias current, $Q = \omega_p R_n C_{\text{JJ}}$ is the junction quality factor, $\Delta U = 2 E_{\text{J0}}(\sqrt{1-\gamma_{\text{JJ}}^2} - \gamma_{\text{JJ}}\cos^{-1}\gamma_{\text{JJ}})$ is the phase particle barrier height and $E_{\text{J0}} = {\hbar I_c}/{2e}$ is the Josephson energy. Here, $e$ is the electron charge. 

\begin{figure*}
\includegraphics[width=1.2\columnwidth]{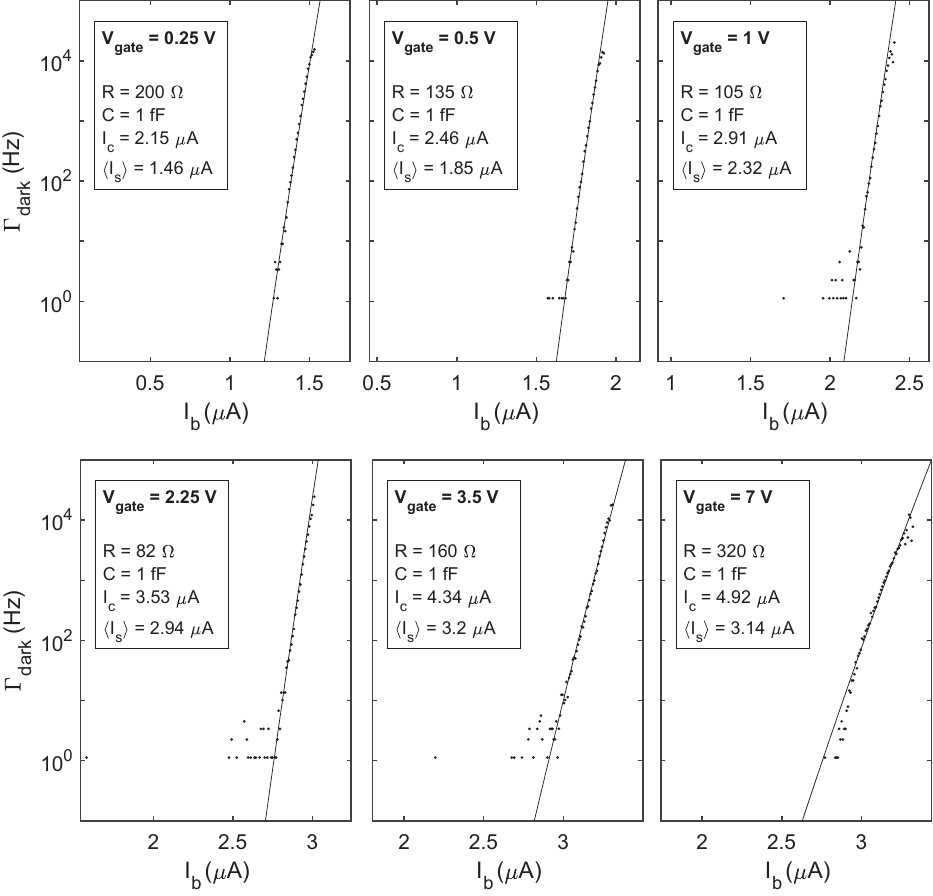}
\caption{\textbf{MQT Fits for the Data Presented in the Main Text.} We use the Fulton-Dunkleberger technique \cite{Fulton.1974} to extract $\Gamma_{\text{dark}}$ for several representative $V_{\text{gate}}$. We then fit using the expressions describing MQT by varying the resistance (R) and $I_c$ for each $V_{\text{gate}}$. }
\label{fig:MQTFits}
\end{figure*}
In Fig. \ref{fig:MQTFits}, we show a fit for each of the $V_{\rm{gate}}$ displayed in Fig. 3A. We find good agreement between the data and fits. Notably, at higher $V_{\rm{gate}}$, the fitted value for resistance begins to increase with increasing $V_{\rm{gate}}$. This trend is not observed in the lock-in measurements of $R_n$. This trend of increasing resistance is further accompanied by an increase in $I_c$ and a qualitative bend in $\Gamma_{\text{meas}}$ versus $I_b$. This may be due to the device matching to a resonance in the electrode wiring, which in turn produces a non-thermal noise as was observed in Ref. \cite{Fulton.1974}. This noise likely inhibits SPD at higher densities. 

In addition to the MQT regime, at higher temperatures the device can enter the thermally activated (TA) regime whereby the thermal fluctuations cause the phase particle to excite over $\Delta$U. In the TA regime, $T_{\text{eff}} = T_e$ and \cite{Devoret.1985}:
\begin{equation}
A = A_{\text{TA}} = \frac{\omega_p}{2\pi}(\sqrt{1+\frac{1}{4Q^2}}-\frac{1}{2Q})
\end{equation}

\subsection{Simple modeling of detection efficiency versus temperature and $\Delta U$}
Fig. 4 shows that the calculated $\eta(T_0,\Delta U/k_B)$ qualitatively agrees with the experimental data. We calculated $\eta$ using \cite{Walsh.2017}:
\begin{equation}
\eta = 1-\exp{ \bigl( -\int \Gamma_{\rm{total}}(t)} \bigr) dt\bigr)
\label{eqn:etaIntegral}
\end{equation} 
where $\Gamma_{\rm{total}}$ is the total switching rate of the GJJ, i.e. $\Gamma_{\rm{total}} = \Gamma_{\rm{MQT}} + \Gamma_{\rm{TA}}$. To induce the phase particle out of the washboard potential, i.e. switching of the GJJ, we can expect the switching rate to follow the general form, $\Gamma = A \exp{\bigl({-\Delta U}/{k_B T_{\text{eff}}}}\bigr)$ based on activation theory. In our devices, we nominally operate the GJJ in the MQT regime such that $\Gamma_{\text{MQT}}$ is dominant. However, when the graphene absorbs a photon, the rise in $T_e$ will increase $\Gamma_{\text{TA}}$, causing a TA switching event. Therefore, achieving a high $\eta$ over the dark count rate requires that the photon to elevate $\Gamma_{\text{TA}} \gg \Gamma_{\rm{MQT}}$. In this case, we can approximate that $T_e = T_{1p}$ so that $\Gamma_{\rm{total}}\simeq \Gamma_{1p} = A \exp{\bigl({-\Delta U}/{k_B T_{1p}}}\bigr)$.

Increasing $T_0$ can degrade calorimetric SPD in two ways. First, increasing ambient thermal fluctuations raises the dark $\Gamma_{\rm{TA}}$ and reduces $I_c$ of the GJJ. However, the $\langle I_s \rangle$ changes only by $\sim$ 30\% over the temperature range studied (a similar change in $\langle I_s \rangle$ is observed when reducing $V_{\text{gate}}$ from 2 V to 1 V). We shall neglect the change of the GJJ's ability to detect a single photon in our model, i.e. the $T_0$ dependence of $A$ and $\Delta U$, as temperature rises.

The second effect, the calorimetric response of the graphene electrons, can dominate the suppression of $\eta$ when $T_0$ rises. Since $T_{1p} = \sqrt{2h\nu/\gamma_s \mathcal{A} + T_0^2}$ \cite{Walsh.2017} with $\mathcal{A}$ being the total area of the monolayer graphene when a uniform $T_e$ is reached \cite{Fried.Fong.2024}, $T_{1p}$ changes considerably when the two terms inside the square root become comparable. Moreover, the integration time in Eqn.~\ref{eqn:etaIntegral} is limited by $\tau_{\rm{ep}}$. Since $\tau_{\text{ep}}(T_0) \propto \tau_{\text{ep}}(T_0 = 20 \text{ mK}) T_0^{2-\delta}$, it reduces quickly as $T_0$ rises. Owing to the relatively large area-to-perimeter ratio of the graphene used in our experiment, it is likely that the E-Ph coupling is in the clean limit ($\delta = 4$) \cite{Hwang.2008,Bistritzer.2009,Betz.2012,Fong.2012,McKitterick.2016,Draelos.2019o4h,Viljas.2010} rather than the disorder or resonant-scattering limit ($\delta = 3$) \cite{Song.2012,Graham.2013,Chen.2012,Fong.2013,Betz.2013,Halbertal.2017}.

To calculate $\eta$, we simplify Eqn. \ref{eqn:etaIntegral} by assuming that the device stays at $\Gamma_{1p}$ for $\tau_{\text{ep}}$:
\begin{equation}
    \eta = 1 - \exp(-\Gamma_{1p}(T_0)\tau_{\text{ep}}(T_0))
    \label{eqn:etaGamma1p}
\end{equation} where we set $A$ in $\Gamma_{1p}$ such that $A = A_{\rm{TA}}(T_0 = 20$~mK).

Using Eqn. \ref{eqn:etaGamma1p}, we vary $\Delta U$ and $T_0$ to calculate $\eta$ (Fig. \ref{fig:Fig4}B). We find qualitative agreement between the experiment and calculation when $\tau_{\text{ep}}(T_0 = 20 \text{mK})$ is 75 ns and $T_{1p}(T_0 = 0)$ is 2.5 K. We note, this model does not capture the behavior at high $\Delta U/k_B$ as it does not account for the retrapping of the junction. However, we see that the overall reduction in $\eta$ for rising $T_0$ is consistent with the experiment. This indicates that our assumption about the unchanging ability of the GJJ to detect a single-photon in this temperature range is reasonable and that the dominant reduction in $\eta$ is due to thermal effects of the graphene electrons.

\subsection{Calculation of the characteristic length scale of heat diffusion}
The characteristic length scale of heat diffusion determines the reduction of $\eta$ at a distance away from the Josephson junction. In the main text we estimate that this length scale to be $\sim$230~$\mu$m. For this we take electrical conductivity $\sigma = 0.02$ S (inferred from the resistance measurement listed in Table~\ref{tab:Devices}) and $\gamma_s = 6.71$ Ws$^{-2}$K$^{-2}$ (corresponding to $n_e = 10^{12}$ cm$^{-2}$). Therefore, $\mathcal{D} = \sigma \mathcal{L}_0 / \gamma_s = 0.727$ m$^2$/s. Taking a characteristic E-Ph interaction time found in Fig. \ref{fig:Fig4} of $\tau_{\text{ep}} = 75$ ns, we find $l_{D} = \sqrt{\mathcal{D}\tau_{\text{ep}}} \simeq $ 230~$\mu$m. As stated in the main text, this length scale is much larger than the device's longitudinal length and likely explained the constant $\eta$ over the device.

\subsection{Behavior of $\eta$ as a function of position from the JJ detector}

The persistence of a high quantum efficiency $\eta$ even when the beam spot is far away from GJJ can be understood through simple modeling of electron heat diffusion in graphene. In our device, electronic heat simultaneously diffuses out through electron-electron interactions and dissipates via collision with the lattice \cite{Massicotte.2021}, as modeled by Eqn.~\ref{eqn:heatdiffeqn2}. Following Ref. \cite{Fried.Fong.2024,Walsh.2017,Aamir.2021}, the value of $\tau_{\text{ep}}$ for different devices can be estimated by $\tau_{\text{ep}} = \gamma_S / \delta \Sigma T_0^{\delta - 2}$. The power law $\delta$ and the strength of E-Ph coupling $\Sigma$ take on different values depending on the mechanism of E-Ph scattering. In the limit of $l_{\text{mfp}}$ larger than the typical inverse phonon momentum, heat dissipation follows a $\delta = 4$ power law, and $\Sigma=\pi^{5/2}k_B^4D^2n_e^{1/2}/(15\rho_m\hbar^4v_F^2s^3)$ \cite{Hwang.2008,Viljas.2010} where $D \simeq 18$~eV is the deformation potential, $\rho_m=7.4\times 10^{-19}$ kg $\mu$m$^{-2}$ is the mass density of graphene, and $s = 2.6\times 10^{4}$ m s$^{-1}$ is the speed of sound in graphene \cite{Sarma.2011}. If $l_{\text{mfp}}$ is lower, heat dissipation is governed by defect-assisted scattering, and follows a $\delta = 3$ power law, with $\Sigma=2\zeta(3)k_B^3D^2n_e^{1/2}/(\pi^{3/2}\rho_m\hbar^3v_F^2s^2l_{\text{mfp}})$ \cite{Song.2012,Chen.2012}. Experimentally, however, a third regime where E-Ph coupling is dominated by resonant scattering on the edge of the graphene is often observed \cite{Lee.2020,Draelos.2019,Halbertal.2017}, especially for samples with a high edge-to-surface ratio. In this regime, heat dissipation also follows a $\delta = 3$ power law, but has a higher value of $\Sigma$, with experimental values close to 1~Wm$^{-2}$K$^{-3}$ \cite{Fong.2013,Draelos.2019,Lee.2020}. While it is difficult to determine with certainty what E-Ph coupling regime our device is in (especially when accounting for the possibility of local differences in scattering across the length of the device), modeling the thermal behavior of a device with our graphene calorimeter's dimensions and parameters in both the diffusive ($\delta$ = 4 clean limit) and dissipative ($\delta$ = 3 resonant scattering) cases can help us understand the range of possibilities we can expect to see for the distance dependence of $\eta$.

First, we determine our initial condition for solving Eqn.~\ref{eqn:heatdiffeqn2}. The absorption of a single near-infrared photon by graphene causes an inter-band electronic excitation, which then results in an initial cascade of hot electron-electron interactions \cite{Massicotte.2021,Song.2013ykg,Brida.2013}. We model this initial heating as a Gaussian hotspot with half-width-half-maximum $\xi$ \cite{Block.2021,Ruzicka.2010}:
\begin{equation}
    T_e(t=0) = T_{\text{hot}} e^{-(x - x_{0})^2/\xi^2} + T_{0}
    \label{eqn:ic}
\end{equation}
where $T_{\text{hot}}$ is the average temperature of the hot electrons in the initial photo-excitation cascade, and $x_0$ is the distance of the hotspot from the center of the graphene. The relationship between the hotspot size $\xi$ and hotspot temperature $T_{\text{hot}}$ can be found through integrating electronic heat capacity $C_e = \mathcal{A}\gamma_S T_e$ with respect to temperature \cite{Walsh.2017,Aamir.2021,Fong.2012,Vora.Du.2012}:
\begin{align}
    \int dE &= \int_{T_0}^{T_{\text{hot}}} \mathcal{A}\gamma_S T_e \, dT_e \label{eqn:hotspot1} \\
    h\nu &= \frac{1}{2} \pi\xi^2\gamma_S \left(T_{\text{hot}}^{2} - T_{0}^{2} \right) \label{eqn:hotspot2}
\end{align}
assuming 100\% integrated electron-electron scattering efficiency \cite{Tielrooij.2013}. There is no definitive formula for $T_{\text{hot}}$ or $\xi$. However, for a 1550 nm incident photon, $T_{\text{hot}} = 100$ K would yield a hotspot size of $\xi =$ 94 nm, that is consistent to the expectation \cite{Block.2021,Song.2013ykg}. 

\begin{figure}[t]\centering
\includegraphics[width=0.8\columnwidth]{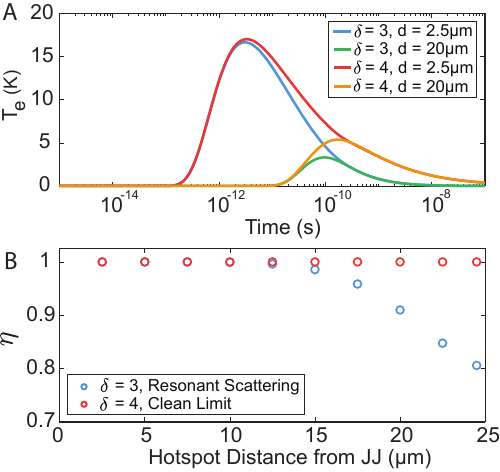}
\caption{\textbf{Numerical simulations of thermal propagation and thermal switching.} Simulations calculated numerically for a graphene calorimetric SPD 25 $\mu$m in length. Panel A shows the time dependent electron temperature measured at different distances from the JJ. Panel B shows the detection efficiency $\eta$ as a function of the hotspot's distance from the Junction. Charge carrier density is set at $2.0 \times 10^{12}$ cm$^{-2}$, and numerical values of the E-Ph coupling constant $\Sigma$ are $0.031 \; $Wm$^{-2}$K$^{-4}$ and $1 \; $Wm$^{-2}$K$^{-3}$ for $\delta = 4$ and $\delta = 3$, respectively.}
\label{fig:Thermal supplement}
\end{figure}

We then choose insulating boundary conditions at $x = 0$ and $x = 25$ $\mu$m in accordance with the dimensions of device B, solving Eqn.~\ref{eqn:heatdiffeqn2} numerically \cite{Fried.Fong.2024} in one dimension for each E-Ph scattering regime. The results are shown in Fig.~\ref{fig:Thermal supplement}A. Depending on the distance between the hotspot and the detector, the timing and magnitude of the initial temperature peak observed will be different. But after a certain amount of time, the entire graphene flake will thermalize to a uniform $T_e$, after which the spatial profile of $T_e$ will be identical regardless of where on the graphene it is measured from. The timescale at which uniform temperature will be reached depends on the size of the device. For a 25 $\mu$m flake of graphene, it is on the order of 100 ps. 

We can take the results of these simulations evaluated at one end of a 25 $\mu$m graphene heterostructure and integrate the time-dependent $T_e$ to calculate the simulated detection efficiency $\eta$ using Eqn. \ref{eqn:etaGamma1p}. Using an $I_c$ of 3.38 $\mu$A and an $I_{b}$ of 2.8 $\mu$A to simulate the conditions seen in Fig. 2, with an integration time of 100 ns, we calculate $\eta$ as a function of hotspot position for the two E-Ph coupling regimes of interest. The results, plotted in Fig. \ref{fig:Thermal supplement}B, show that $\eta$ may decrease with increasing distance away from the detector for $\delta = 3$. However, for $\delta = 4$, there is no appreciable reduction in $\eta$ over 25 $\mu$m. The reason for this lies in the temperature at which the graphene reaches a uniform distribution of electronic heat. In the $\delta = 4$ case, after the graphene reaches a uniform temperature, it remains at a temperature substantially above $\Delta U/k_{B}$ for a sufficiently long amount of time that, regardless of the size of the initial peak during the first 0.1 ns, the junction will have 100\% probability of switching. In contrast, the uniform temperature reached in the $\delta = 3$ case is lower, and remains elevated for less time, thermalizing to base temperature before 10 ns have elapsed. As such, whether the detector experiences elevated temperatures during the first 0.1 ns after photon absorption has a significant effect on the detection efficiency, causing a distance dependence in $\eta$.

1-D thermal modeling is not sufficient to definitively show that the devices measured in this paper follow the E-Ph coupling characteristic of the clean regime ($\delta = 4$) --- the presence of multiple different E-Ph coupling mechanisms affecting the graphene cannot be ruled out. However, these calculations demonstrate that the independence of $\eta$ as a function of distance from the JJ as seen in our device is consistent with a realistic model of heat propagation in graphene.

\clearpage

\begin{thebibliography}{10}
\expandafter\ifx\csname url\endcsname\relax
  \def\url#1{\texttt{#1}}\fi
\expandafter\ifx\csname urlprefix\endcsname\relax\def\urlprefix{URL }\fi
\providecommand{\bibinfo}[2]{#2}
\providecommand{\eprint}[2][]{\url{#2}}

\bibitem{Hadfield.2009}
\bibinfo{author}{Hadfield, R.~H.}
\newblock \bibinfo{title}{{Single-photon detectors for optical quantum information applications}}.
\newblock \emph{\bibinfo{journal}{Nature Photonics}} \textbf{\bibinfo{volume}{3}}, \bibinfo{pages}{696} (\bibinfo{year}{2009}).
\newblock \urlprefix\url{http://www.nature.com/nphoton/journal/v3/n12/abs/nphoton.2009.230.html}.

\bibitem{Eisaman.2011}
\bibinfo{author}{Eisaman, M.~D.}, \bibinfo{author}{Fan, J.}, \bibinfo{author}{Migdall, A.} \& \bibinfo{author}{Polyakov, S.~V.}
\newblock \bibinfo{title}{{Invited Review Article: Single-photon sources and detectors}}.
\newblock \emph{\bibinfo{journal}{Review Of Scientific Instruments}} \textbf{\bibinfo{volume}{82}}, \bibinfo{pages}{071101} (\bibinfo{year}{2011}).
\newblock \urlprefix\url{http://aip.scitation.org/doi/10.1063/1.3610677}.

\bibitem{Couteau.2023}
\bibinfo{author}{Couteau, C.} \emph{et~al.}
\newblock \bibinfo{title}{{Applications of single photons to quantum communication and computing}}.
\newblock \emph{\bibinfo{journal}{Nature Reviews Physics}} \textbf{\bibinfo{volume}{5}}, \bibinfo{pages}{326--338} (\bibinfo{year}{2023}).

\bibitem{Bruschini.Charbon.2019}
\bibinfo{author}{Bruschini, C.}, \bibinfo{author}{Homulle, H.}, \bibinfo{author}{Antolovic, I.~M.}, \bibinfo{author}{Burri, S.} \& \bibinfo{author}{Charbon, E.}
\newblock \bibinfo{title}{{Single-photon avalanche diode imagers in biophotonics: review and outlook}}.
\newblock \emph{\bibinfo{journal}{Light: Science \& Applications}} \textbf{\bibinfo{volume}{8}}, \bibinfo{pages}{87} (\bibinfo{year}{2019}).

\bibitem{Kirmani.2014}
\bibinfo{author}{Kirmani, A.} \emph{et~al.}
\newblock \bibinfo{title}{{First-Photon Imaging}}.
\newblock \emph{\bibinfo{journal}{Science}} \textbf{\bibinfo{volume}{343}}, \bibinfo{pages}{58 -- 61} (\bibinfo{year}{2014}).
\newblock \urlprefix\url{https://www.sciencemag.org/lookup/doi/10.1126/science.1246775}.

\bibitem{Ceccarelli.Osellame.2021}
\bibinfo{author}{Ceccarelli, F.} \emph{et~al.}
\newblock \bibinfo{title}{{Recent Advances and Future Perspectives of Single‐Photon Avalanche Diodes for Quantum Photonics Applications}}.
\newblock \emph{\bibinfo{journal}{Advanced Quantum Technologies}} \textbf{\bibinfo{volume}{4}} (\bibinfo{year}{2021}).
\newblock \eprint{2010.05613}.

\bibitem{Seifert.2020}
\bibinfo{author}{Seifert, P.} \emph{et~al.}
\newblock \bibinfo{title}{{Magic-Angle Bilayer Graphene Nanocalorimeters: Toward Broadband, Energy-Resolving Single Photon Detection}}.
\newblock \emph{\bibinfo{journal}{Nano Letters}} \textbf{\bibinfo{volume}{20}}, \bibinfo{pages}{3459} (\bibinfo{year}{2020}).
\newblock \urlprefix\url{https://pubs.acs.org/doi/10.1021/acs.nanolett.0c00373}.

\bibitem{Verma.2021}
\bibinfo{author}{Verma, V.~B.} \emph{et~al.}
\newblock \bibinfo{title}{{Single-photon detection in the mid-infrared up to 10 um wavelength using tungsten silicide superconducting nanowire detectors}}.
\newblock \emph{\bibinfo{journal}{APL Photonics}} \textbf{\bibinfo{volume}{6}}, \bibinfo{pages}{056101} (\bibinfo{year}{2021}).

\bibitem{Fong.2012}
\bibinfo{author}{Fong, K.~C.} \& \bibinfo{author}{Schwab, K.}
\newblock \bibinfo{title}{{Ultrasensitive and wide-bandwidth thermal measurements of graphene at low temperatures}}.
\newblock \emph{\bibinfo{journal}{Physical Review X}} \textbf{\bibinfo{volume}{2}}, \bibinfo{pages}{031006} (\bibinfo{year}{2012}).

\bibitem{Du.2014}
\bibinfo{author}{Du, X.}, \bibinfo{author}{Prober, D.~E.}, \bibinfo{author}{Vora, H.} \& \bibinfo{author}{Mckitterick, C.~B.}
\newblock \bibinfo{title}{{Graphene-based Bolometers}}.
\newblock \emph{\bibinfo{journal}{Graphene and 2D Materials}} \textbf{\bibinfo{volume}{1}} (\bibinfo{year}{2014}).

\bibitem{Walsh.2017}
\bibinfo{author}{Walsh, E.~D.} \emph{et~al.}
\newblock \bibinfo{title}{{Graphene-Based Josephson-Junction Single-Photon Detector}}.
\newblock \emph{\bibinfo{journal}{Physical Review Applied}} \textbf{\bibinfo{volume}{8}}, \bibinfo{pages}{024022} (\bibinfo{year}{2017}).
\newblock \urlprefix\url{https://link.aps.org/doi/10.1103/PhysRevApplied.8.024022}.

\bibitem{Lee.2020}
\bibinfo{author}{Lee, G.-H.} \emph{et~al.}
\newblock \bibinfo{title}{{Graphene-based Josephson junction microwave bolometer}}.
\newblock \emph{\bibinfo{journal}{Nature}} \textbf{\bibinfo{volume}{586}}, \bibinfo{pages}{42 -- 46} (\bibinfo{year}{2020}).
\newblock \urlprefix\url{https://www.nature.com/articles/s41586-020-2752-4}.

\bibitem{Kokkoniemi.2020}
\bibinfo{author}{Kokkoniemi, R.} \emph{et~al.}
\newblock \bibinfo{title}{{Bolometer operating at the threshold for circuit quantum electrodynamics}}.
\newblock \emph{\bibinfo{journal}{Nature}} \textbf{\bibinfo{volume}{586}}, \bibinfo{pages}{47 -- 51} (\bibinfo{year}{2020}).
\newblock \urlprefix\url{https://www.nature.com/articles/s41586-020-2753-3}.

\bibitem{Sarma.2011}
\bibinfo{author}{Sarma, S.~D.}, \bibinfo{author}{Adam, S.}, \bibinfo{author}{Hwang, E.~H.} \& \bibinfo{author}{Rossi, E.}
\newblock \bibinfo{title}{{Electronic transport in two-dimensional graphene}}.
\newblock \emph{\bibinfo{journal}{Reviews Of Modern Physics}} \textbf{\bibinfo{volume}{83}}, \bibinfo{pages}{407 -- 470} (\bibinfo{year}{2011}).
\newblock \urlprefix\url{https://link.aps.org/doi/10.1103/RevModPhys.83.407}.

\bibitem{Echternach.2018}
\bibinfo{author}{Echternach, P.}, \bibinfo{author}{Pepper, B.}, \bibinfo{author}{Reck, T.} \& \bibinfo{author}{Bradford, C.}
\newblock \bibinfo{title}{Single photon detection of 1.5 thz radiation with the quantum capacitance detector}.
\newblock \emph{\bibinfo{journal}{Nature Astronomy}} \textbf{\bibinfo{volume}{2}}, \bibinfo{pages}{90–97} (\bibinfo{year}{2017}).
\newblock \urlprefix\url{http://dx.doi.org/10.1038/s41550-017-0294-y}.

\bibitem{day.2024}
\bibinfo{author}{Day, P.~K.} \emph{et~al.}
\newblock \bibinfo{title}{A 25-micrometer single-photon-sensitive kinetic inductance detector}.
\newblock \emph{\bibinfo{journal}{Physical Review X}} \textbf{\bibinfo{volume}{14}} (\bibinfo{year}{2024}).
\newblock \urlprefix\url{http://dx.doi.org/10.1103/physrevx.14.041005}.

\bibitem{Hochberg.2019}
\bibinfo{author}{Hochberg, Y.} \emph{et~al.}
\newblock \bibinfo{title}{{Detecting Sub-GeV Dark Matter with Superconducting Nanowires}}.
\newblock \emph{\bibinfo{journal}{Physical Review Letters}} \textbf{\bibinfo{volume}{123}}, \bibinfo{pages}{151802} (\bibinfo{year}{2019}).
\newblock \urlprefix\url{https://link.aps.org/doi/10.1103/PhysRevLett.123.151802}.

\bibitem{Dixit.2021}
\bibinfo{author}{Dixit, A.~V.} \emph{et~al.}
\newblock \bibinfo{title}{{Searching for Dark Matter with a Superconducting Qubit}}.
\newblock \emph{\bibinfo{journal}{Physical Review Letters}} \textbf{\bibinfo{volume}{126}}, \bibinfo{pages}{141302} (\bibinfo{year}{2021}).
\newblock \eprint{2008.12231}.

\bibitem{Lauk.2020}
\bibinfo{author}{Lauk, N.} \emph{et~al.}
\newblock \bibinfo{title}{{Perspectives on quantum transduction}}.
\newblock \emph{\bibinfo{journal}{Quantum Science and Technology}} \textbf{\bibinfo{volume}{5}}, \bibinfo{pages}{020501} (\bibinfo{year}{2020}).
\newblock \urlprefix\url{https://iopscience.iop.org/article/10.1088/2058-9565/ab788a}.

\bibitem{Gol'tsman.2001}
\bibinfo{author}{Gol'tsman, G.~N.} \emph{et~al.}
\newblock \bibinfo{title}{{Picosecond superconducting single-photon optical detector}}.
\newblock \emph{\bibinfo{journal}{Applied Physics Letters}} \textbf{\bibinfo{volume}{79}}, \bibinfo{pages}{705 -- 707} (\bibinfo{year}{2001}).
\newblock \urlprefix\url{http://aip.scitation.org/doi/10.1063/1.1388868}.

\bibitem{Charaev.2023}
\bibinfo{author}{Charaev, I.} \emph{et~al.}
\newblock \bibinfo{title}{{Single-photon detection using high-temperature superconductors}}.
\newblock \emph{\bibinfo{journal}{Nature Nanotechnology}} \textbf{\bibinfo{volume}{18}}, \bibinfo{pages}{343--349} (\bibinfo{year}{2023}).
\newblock \eprint{2208.05674}.

\bibitem{Ullom.Bennett.2015}
\bibinfo{author}{Ullom, J.~N.} \& \bibinfo{author}{Bennett, D.~A.}
\newblock \bibinfo{title}{{Review of superconducting transition-edge sensors for x-ray and gamma-ray spectroscopy}}.
\newblock \emph{\bibinfo{journal}{Superconductor Science and Technology}} \textbf{\bibinfo{volume}{28}}, \bibinfo{pages}{084003} (\bibinfo{year}{2015}).

\bibitem{Hwang.2008}
\bibinfo{author}{Hwang, E.~H.} \& \bibinfo{author}{Sarma, S.~D.}
\newblock \bibinfo{title}{{Acoustic phonon scattering limited carrier mobility in two-dimensional extrinsic graphene}}.
\newblock \emph{\bibinfo{journal}{Physical Review B}} \textbf{\bibinfo{volume}{77}}, \bibinfo{pages}{115449} (\bibinfo{year}{2008}).
\newblock \urlprefix\url{https://link.aps.org/doi/10.1103/PhysRevB.77.115449}.

\bibitem{song.levitov.2015}
\bibinfo{author}{Song, J. C.~W.} \& \bibinfo{author}{Levitov, L.~S.}
\newblock \bibinfo{title}{{Energy flows in graphene: hot carrier dynamics and cooling}}.
\newblock \emph{\bibinfo{journal}{Journal of Physics: Condensed Matter}} \textbf{\bibinfo{volume}{27}}, \bibinfo{pages}{164201} (\bibinfo{year}{2015}).
\newblock \eprint{1410.5426}.

\bibitem{Zgirski.2018}
\bibinfo{author}{Zgirski, M.} \emph{et~al.}
\newblock \bibinfo{title}{{Nanosecond Thermometry with Josephson Junctions}}.
\newblock \emph{\bibinfo{journal}{Physical Review Applied}} \textbf{\bibinfo{volume}{10}}, \bibinfo{pages}{044068} (\bibinfo{year}{2018}).
\newblock \eprint{1704.04762}.

\bibitem{Tielrooij.2013}
\bibinfo{author}{Tielrooij, K.~J.} \emph{et~al.}
\newblock \bibinfo{title}{{Photoexcitation cascade and multiple hot-carrier generation in graphene}}.
\newblock \emph{\bibinfo{journal}{Nature Physics}} \textbf{\bibinfo{volume}{9}}, \bibinfo{pages}{248} (\bibinfo{year}{2013}).
\newblock \urlprefix\url{http://www.nature.com.ezproxy.cul.columbia.edu/nphys/journal/v9/n4/full/nphys2564.html}.

\bibitem{Block.2021}
\bibinfo{author}{Block, A.} \emph{et~al.}
\newblock \bibinfo{title}{{Observation of giant and tunable thermal diffusivity of a Dirac fluid at room temperature}}.
\newblock \emph{\bibinfo{journal}{Nature Nanotechnology}} \textbf{\bibinfo{volume}{16}}, \bibinfo{pages}{1195--1200} (\bibinfo{year}{2021}).
\newblock \eprint{2008.04189}.

\bibitem{Tinkham}
\bibinfo{author}{Tinkham, M.}
\newblock \emph{\bibinfo{title}{{Introduction to Superconductivity}}} (\bibinfo{publisher}{McGraw-Hill}, \bibinfo{year}{1996}).

\bibitem{Martinis.1987}
\bibinfo{author}{Martinis, J.~M.}, \bibinfo{author}{Devoret, M.~H.} \& \bibinfo{author}{Clarke, J.}
\newblock \bibinfo{title}{{Experimental tests for the quantum behavior of a macroscopic degree of freedom: The phase difference across a Josephson junction}}.
\newblock \emph{\bibinfo{journal}{Physical Review B}} \textbf{\bibinfo{volume}{35}}, \bibinfo{pages}{4682--4698} (\bibinfo{year}{1987}).
\newblock \urlprefix\url{http://gateway.webofknowledge.com/gateway/Gateway.cgi?GWVersion=2\&SrcAuth=mekentosj\&SrcApp=Papers\&DestLinkType=FullRecord\&DestApp=WOS\&KeyUT=A1987G717000018}.

\bibitem{Devoret.1985}
\bibinfo{author}{Devoret, M.~H.}, \bibinfo{author}{Martinis, J.~M.} \& \bibinfo{author}{Clarke, J.}
\newblock \bibinfo{title}{{Measurements of Macroscopic Quantum Tunneling out of the Zero-Voltage State of a Current-Biased Josephson Junction}}.
\newblock \emph{\bibinfo{journal}{Physical Review Letters}} \textbf{\bibinfo{volume}{55}}, \bibinfo{pages}{1908 -- 1911} (\bibinfo{year}{1985}).
\newblock \urlprefix\url{http://link.aps.org/doi/10.1103/PhysRevLett.55.1908}.

\bibitem{Courtois.2008}
\bibinfo{author}{Courtois, H.}, \bibinfo{author}{Meschke, M.}, \bibinfo{author}{Peltonen, J.~T.} \& \bibinfo{author}{Pekola, J.~P.}
\newblock \bibinfo{title}{{Origin of Hysteresis in a Proximity Josephson Junction}}.
\newblock \emph{\bibinfo{journal}{Physical Review Letters}} \textbf{\bibinfo{volume}{101}}, \bibinfo{pages}{067002} (\bibinfo{year}{2008}).
\newblock \urlprefix\url{http://link.aps.org/doi/10.1103/PhysRevLett.101.067002}.

\bibitem{Borzenets.2016nj}
\bibinfo{author}{Borzenets, I.~V.} \emph{et~al.}
\newblock \bibinfo{title}{{Ballistic Graphene Josephson Junctions from the Short to the Long Junction Regimes}}.
\newblock \emph{\bibinfo{journal}{Physical Review Letters}} \textbf{\bibinfo{volume}{117}}, \bibinfo{pages}{237002} (\bibinfo{year}{2016}).
\newblock \urlprefix\url{http://link.aps.org/doi/10.1103/PhysRevLett.117.237002}.

\bibitem{Walsh.2021}
\bibinfo{author}{Walsh, E.~D.} \emph{et~al.}
\newblock \bibinfo{title}{{Josephson junction infrared single-photon detector}}.
\newblock \emph{\bibinfo{journal}{Science}} \textbf{\bibinfo{volume}{372}}, \bibinfo{pages}{409 -- 412} (\bibinfo{year}{2021}).
\newblock \urlprefix\url{https://science.sciencemag.org/content/372/6540/409}.

\bibitem{si}
\bibinfo{title}{{See Supplementary Information for details.}}

\bibitem{Gan.2012}
\bibinfo{author}{Gan, X.} \emph{et~al.}
\newblock \bibinfo{title}{{Strong Enhancement of Light–Matter Interaction in Graphene Coupled to a Photonic Crystal Nanocavity}}.
\newblock \emph{\bibinfo{journal}{Nano Letters}} \textbf{\bibinfo{volume}{12}}, \bibinfo{pages}{5626 -- 5631} (\bibinfo{year}{2012}).
\newblock \urlprefix\url{http://pubs.acs.org/doi/abs/10.1021/nl302746n}.

\bibitem{Furchi.Mueller.2012}
\bibinfo{author}{Furchi, M.} \emph{et~al.}
\newblock \bibinfo{title}{{Microcavity-Integrated Graphene Photodetector}}.
\newblock \emph{\bibinfo{journal}{Nano Letters}} \textbf{\bibinfo{volume}{12}}, \bibinfo{pages}{2773--2777} (\bibinfo{year}{2012}).
\newblock \eprint{1112.1549}.

\bibitem{Vasic.2014}
\bibinfo{author}{Vasić, B.} \& \bibinfo{author}{Gajić, R.}
\newblock \bibinfo{title}{{Tunable Fabry–Perot resonators with embedded graphene from terahertz to near-infrared frequencies}}.
\newblock \emph{\bibinfo{journal}{Optics Letters}} \textbf{\bibinfo{volume}{39}}, \bibinfo{pages}{6253} (\bibinfo{year}{2014}).

\bibitem{Fried.Fong.2024}
\bibinfo{author}{Fried, C.} \emph{et~al.}
\newblock \bibinfo{title}{{Performance limits due to thermal transport in graphene single-photon bolometers}}.
\newblock \emph{\bibinfo{journal}{Physical Review Applied}} \textbf{\bibinfo{volume}{21}}, \bibinfo{pages}{014006} (\bibinfo{year}{2024}).
\newblock \eprint{2311.00228}.

\bibitem{Aumentado.2004}
\bibinfo{author}{Aumentado, J.}, \bibinfo{author}{Keller, M.~W.}, \bibinfo{author}{Martinis, J.~M.} \& \bibinfo{author}{Devoret, M.~H.}
\newblock \bibinfo{title}{{Nonequilibrium Quasiparticles and 2e Periodicity in Single-Cooper-Pair Transistors}}.
\newblock \emph{\bibinfo{journal}{Physical Review Letters}} \textbf{\bibinfo{volume}{92}}, \bibinfo{pages}{066802} (\bibinfo{year}{2004}).
\newblock \eprint{cond-mat/0308253}.

\bibitem{Lee.2011}
\bibinfo{author}{Lee, G.-H.}, \bibinfo{author}{Jeong, D.}, \bibinfo{author}{Choi, J.-H.}, \bibinfo{author}{Doh, Y.-J.} \& \bibinfo{author}{Lee, H.-J.}
\newblock \bibinfo{title}{{Electrically Tunable Macroscopic Quantum Tunneling in a Graphene-Based Josephson Junction}}.
\newblock \emph{\bibinfo{journal}{Physical Review Letters}} \textbf{\bibinfo{volume}{107}}, \bibinfo{pages}{146605} (\bibinfo{year}{2011}).
\newblock \urlprefix\url{http://link.aps.org/doi/10.1103/PhysRevLett.107.146605}.

\bibitem{Fulton.1974}
\bibinfo{author}{Fulton, T.~A.} \& \bibinfo{author}{Dunkleberger, L.~N.}
\newblock \bibinfo{title}{{Lifetime of the zero-voltage state in Josephson tunnel junctions}}.
\newblock \emph{\bibinfo{journal}{Physical Review B}} \textbf{\bibinfo{volume}{9}}, \bibinfo{pages}{4760--4768} (\bibinfo{year}{1974}).
\newblock \urlprefix\url{http://gateway.webofknowledge.com/gateway/Gateway.cgi?GWVersion=2\&SrcAuth=mekentosj\&SrcApp=Papers\&DestLinkType=FullRecord\&DestApp=WOS\&KeyUT=A1974T236000017}.

\bibitem{Giazotto.2008}
\bibinfo{author}{Giazotto, F.} \emph{et~al.}
\newblock \bibinfo{title}{{Ultrasensitive proximity Josephson sensor with kinetic inductance readout}}.
\newblock \emph{\bibinfo{journal}{Applied Physics Letters}} \textbf{\bibinfo{volume}{92}}, \bibinfo{pages}{162507} (\bibinfo{year}{2008}).
\newblock \urlprefix\url{http://scitation.aip.org/content/aip/journal/apl/92/16/10.1063/1.2908922}.

\bibitem{Katti.2022}
\bibinfo{author}{Katti, R.} \emph{et~al.}
\newblock \bibinfo{title}{{Hot Carrier Thermalization and Josephson Inductance Thermometry in a Graphene-based Microwave Circuit}}.
\newblock \emph{\bibinfo{journal}{arXiv}}  (\bibinfo{year}{2022}).
\newblock \eprint{2208.13379}.

\bibitem{Wang.2013}
\bibinfo{author}{Wang, L.} \emph{et~al.}
\newblock \bibinfo{title}{{One-Dimensional Electrical Contact to a Two-Dimensional Material}}.
\newblock \emph{\bibinfo{journal}{Science}} \textbf{\bibinfo{volume}{342}}, \bibinfo{pages}{614 -- 617} (\bibinfo{year}{2013}).
\newblock \urlprefix\url{http://www.sciencemag.org/cgi/doi/10.1126/science.1244358}.

\bibitem{Walsh.2020}
\bibinfo{author}{Walsh, E.~D.}
\newblock \emph{\bibinfo{title}{{Detecting Single Photons with Graphene-Based Josephson Junctions}}}.
\newblock Ph.D. thesis (\bibinfo{year}{2020}).

\bibitem[S46]{Heersche.2007}
\bibinfo{author}{Heersche, H.~B.}, \bibinfo{author}{Jarillo-Herrero, P.}, \bibinfo{author}{Oostinga, J.~B.}, \bibinfo{author}{Vandersypen, L. M.~K.} \& \bibinfo{author}{Morpurgo, A.~F.}
\newblock \bibinfo{title}{{Bipolar supercurrent in graphene}}.
\newblock \emph{\bibinfo{journal}{Nature}} \textbf{\bibinfo{volume}{446}}, \bibinfo{pages}{56} (\bibinfo{year}{2007}).
\newblock \urlprefix\url{http://www.nature.com/nature/journal/v446/n7131/full/nature05555.html}.

\bibitem[S47]{Du.2008}
\bibinfo{author}{Du, X.}, \bibinfo{author}{Skachko, I.} \& \bibinfo{author}{Andrei, E.~Y.}
\newblock \bibinfo{title}{{Josephson current and multiple Andreev reflections in graphene SNS junctions}}.
\newblock \emph{\bibinfo{journal}{Physical Review B}} \textbf{\bibinfo{volume}{77}} (\bibinfo{year}{2008}).
\newblock \urlprefix\url{http://link.aps.org/doi/10.1103/PhysRevB.77.184507}.
\newblock \eprint{0710.4984}.

\bibitem[S48]{Shalom.2015}
\bibinfo{author}{Shalom, M.~B.} \emph{et~al.}
\newblock \bibinfo{title}{{Quantum oscillations of the critical current and high-field superconducting proximity in ballistic graphene}}.
\newblock \emph{\bibinfo{journal}{Nature Physics}} \textbf{\bibinfo{volume}{12}}, \bibinfo{pages}{318 -- 322} (\bibinfo{year}{2015}).
\newblock \urlprefix\url{http://www.nature.com/doifinder/10.1038/nphys3592}.

\bibitem[S49]{Calado.2015}
\bibinfo{author}{Calado, V.~E.} \emph{et~al.}
\newblock \bibinfo{title}{{Ballistic Josephson junctions in edge-contacted graphene}}.
\newblock \emph{\bibinfo{journal}{Nature Nanotechnology}} \textbf{\bibinfo{volume}{10}}, \bibinfo{pages}{761 -- 764} (\bibinfo{year}{2015}).
\newblock \urlprefix\url{http://www.nature.com/doifinder/10.1038/nnano.2015.156}.

\bibitem[S50]{Holzman.2019}
\bibinfo{author}{Holzman, I.} \& \bibinfo{author}{Ivry, Y.}
\newblock \bibinfo{title}{{Superconducting Nanowires for Single‐Photon Detection: Progress, Challenges, and Opportunities}}.
\newblock \emph{\bibinfo{journal}{Advanced Quantum Technologies}} \textbf{\bibinfo{volume}{2}} (\bibinfo{year}{2019}).
\newblock \eprint{1807.09060}.

\bibitem[S51]{Gaggero.2010}
\bibinfo{author}{Gaggero, A.} \emph{et~al.}
\newblock \bibinfo{title}{{Nanowire superconducting single-photon detectors on GaAs for integrated quantum photonic applications}}.
\newblock \emph{\bibinfo{journal}{Applied Physics Letters}} \textbf{\bibinfo{volume}{97}}, \bibinfo{pages}{151108} (\bibinfo{year}{2010}).

\bibitem[S52]{Reddy.2020}
\bibinfo{author}{Reddy, D.~V.}, \bibinfo{author}{Nerem, R.~R.}, \bibinfo{author}{Nam, S.~W.}, \bibinfo{author}{Mirin, R.~P.} \& \bibinfo{author}{Verma, V.~B.}
\newblock \bibinfo{title}{{Superconducting nanowire single-photon detectors with 98\% system detection efficiency at 1550 nm}}.
\newblock \emph{\bibinfo{journal}{Optica}} \textbf{\bibinfo{volume}{7}}, \bibinfo{pages}{1649} (\bibinfo{year}{2020}).

\bibitem[S53]{Korzh.2020}
\bibinfo{author}{Korzh, B.} \emph{et~al.}
\newblock \bibinfo{title}{{Demonstration of sub-3 ps temporal resolution with a superconducting nanowire single-photon detector}}.
\newblock \emph{\bibinfo{journal}{Nature Photonics}} \textbf{\bibinfo{volume}{14}}, \bibinfo{pages}{250 -- 255} (\bibinfo{year}{2020}).
\newblock \urlprefix\url{https://www.nature.com/articles/s41566-020-0589-x}.

\bibitem[S54]{Irwin.Hilton.2005}
\bibinfo{author}{Irwin, K.} \& \bibinfo{author}{Hilton, G.}
\newblock \bibinfo{title}{{Cryogenic Particle Detection}}.
\newblock \emph{\bibinfo{journal}{Topics in Applied Physics}} \bibinfo{pages}{63--150} (\bibinfo{year}{2005}).

\bibitem[S55]{Patel.Thomas.2024}
\bibinfo{author}{Patel, K.~M.}, \bibinfo{author}{Withington, S.}, \bibinfo{author}{Shard, A. G.~.}, \bibinfo{author}{Goldie, D.~J.} \& \bibinfo{author}{Thomas, C.~N.}
\newblock \bibinfo{title}{{Electron spectroscopy using transition-edge sensors}}.
\newblock \emph{\bibinfo{journal}{Journal of Applied Physics}} \textbf{\bibinfo{volume}{135}}, \bibinfo{pages}{224504} (\bibinfo{year}{2024}).

\bibitem[S56]{Efetov.2018}
\bibinfo{author}{Efetov, D.~K.} \emph{et~al.}
\newblock \bibinfo{title}{{Fast thermal relaxation in cavity-coupled graphene bolometers with a Johnson noise read-out}}.
\newblock \emph{\bibinfo{journal}{Nature Nanotechnology}} \textbf{\bibinfo{volume}{13}}, \bibinfo{pages}{797 -- 801} (\bibinfo{year}{2018}).
\newblock \urlprefix\url{https://www.nature.com/articles/s41565-018-0169-0}.

\bibitem[S57]{Deng.2015}
\bibinfo{author}{Deng, X.-H.}, \bibinfo{author}{Liu, J.-T.}, \bibinfo{author}{Yuan, J.-R.}, \bibinfo{author}{Liao, Q.-H.} \& \bibinfo{author}{Liu, N.-H.}
\newblock \bibinfo{title}{A new transfer matrix method to calculate the optical absorption of graphene at any position in stratified media}.
\newblock \emph{\bibinfo{journal}{Europhysics Letters}} \textbf{\bibinfo{volume}{109}}, \bibinfo{pages}{27002} (\bibinfo{year}{2015}).
\newblock \urlprefix\url{https://dx.doi.org/10.1209/0295-5075/109/27002}.

\bibitem[S58]{SensaleRodriguez.2012}
\bibinfo{author}{Sensale-Rodriguez, B.} \emph{et~al.}
\newblock \bibinfo{title}{Broadband graphene terahertz modulators enabled by intraband transitions}.
\newblock \emph{\bibinfo{journal}{Nature Communications}} \textbf{\bibinfo{volume}{3}}, \bibinfo{pages}{780} (\bibinfo{year}{2012}).
\newblock \urlprefix\url{https://doi.org/10.1038/ncomms1787}.

\bibitem[S59]{Mehew.2024}
\bibinfo{author}{Mehew, J.~D.} \emph{et~al.}
\newblock \bibinfo{title}{Ultrafast umklapp-assisted electron-phonon cooling in magic-angle twisted bilayer graphene}.
\newblock \emph{\bibinfo{journal}{Science Advances}} \textbf{\bibinfo{volume}{10}}, \bibinfo{pages}{eadj1361} (\bibinfo{year}{2024}).
\newblock \urlprefix\url{https://www.science.org/doi/abs/10.1126/sciadv.adj1361}.
\newblock \eprint{https://www.science.org/doi/pdf/10.1126/sciadv.adj1361}.

\bibitem[S60]{Saleh.2019}
\bibinfo{author}{Saleh, B.} \& \bibinfo{author}{Teich, M.}
\newblock \emph{\bibinfo{title}{{Fundamentals of Photonics}}}, vol.~\bibinfo{volume}{1} of \emph{\bibinfo{series}{Wiley Series in Pure and Applied Optics}}.

\bibitem[S61]{Marom.1979}
\bibinfo{author}{Marom, E.}, \bibinfo{author}{Chen, B.} \& \bibinfo{author}{Ramer, O.~G.}
\newblock \bibinfo{title}{{Spot Size of Focused Truncated Gaussian Beams}}.
\newblock \emph{\bibinfo{journal}{Optical Engineering}} \textbf{\bibinfo{volume}{18}}, \bibinfo{pages}{180179} (\bibinfo{year}{1979}).
\newblock \urlprefix\url{https://doi.org/10.1117/12.7972325}.

\bibitem[S62]{Urey.2004}
\bibinfo{author}{Urey, H.}
\newblock \bibinfo{title}{Spot size, depth-of-focus, and diffraction ring intensity formulas for truncated gaussian beams}.
\newblock \emph{\bibinfo{journal}{Appl. Opt.}} \textbf{\bibinfo{volume}{43}}, \bibinfo{pages}{620--625} (\bibinfo{year}{2004}).
\newblock \urlprefix\url{https://opg.optica.org/ao/abstract.cfm?URI=ao-43-3-620}.

\bibitem[S63]{Yan.Fuhrer.2012}
\bibinfo{author}{Yan, J.} \emph{et~al.}
\newblock \bibinfo{title}{{Dual-gated bilayer graphene hot-electron bolometer}}.
\newblock \emph{\bibinfo{journal}{Nature Nanotechnology}} \textbf{\bibinfo{volume}{7}}, \bibinfo{pages}{472} (\bibinfo{year}{2012}).
\newblock \urlprefix\url{http://www.nature.com/nnano/journal/v7/n7/full/nnano.2012.88.html}.

\bibitem[S64]{Vora.Du.2012}
\bibinfo{author}{Vora, H.}, \bibinfo{author}{Kumaravadivel, P.}, \bibinfo{author}{Nielsen, B.} \& \bibinfo{author}{Du, X.}
\newblock \bibinfo{title}{{Bolometric response in graphene based superconducting tunnel junctions}}.
\newblock \emph{\bibinfo{journal}{Applied Physics Letters}} \textbf{\bibinfo{volume}{100}}, \bibinfo{pages}{153507} (\bibinfo{year}{2012}).
\newblock \urlprefix\url{http://scitation.aip.org/content/aip/journal/apl/100/15/10.1063/1.3703117}.

\bibitem[S65]{Fatimy.2016}
\bibinfo{author}{Fatimy, A.~E.} \emph{et~al.}
\newblock \bibinfo{title}{{Epitaxial graphene quantum dots for high-performance terahertz bolometers}}.
\newblock \emph{\bibinfo{journal}{Nature Nanotechnology}} \textbf{\bibinfo{volume}{11}}, \bibinfo{pages}{335 -- 338} (\bibinfo{year}{2016}).
\newblock \urlprefix\url{https://www.nature.com/articles/nnano.2015.303}.

\bibitem[S66]{Cai.2014}
\bibinfo{author}{Cai, X.} \emph{et~al.}
\newblock \bibinfo{title}{{Sensitive room-temperature terahertz detection via the photothermoelectric effect in graphene}}.
\newblock \emph{\bibinfo{journal}{Nature Nanotechnology}} \textbf{\bibinfo{volume}{9}}, \bibinfo{pages}{814 -- 819} (\bibinfo{year}{2014}).
\newblock \urlprefix\url{https://www.nature.com/articles/nnano.2014.182}.

\bibitem[S67]{Blaikie.2019}
\bibinfo{author}{Blaikie, A.}, \bibinfo{author}{Miller, D.} \& \bibinfo{author}{Alemán, B.~J.}
\newblock \bibinfo{title}{{A fast and sensitive room-temperature graphene nanomechanical bolometer}}.
\newblock \emph{\bibinfo{journal}{Nature Communications}} \textbf{\bibinfo{volume}{10}}, \bibinfo{pages}{1 -- 8} (\bibinfo{year}{2019}).
\newblock \urlprefix\url{https://www.nature.com/articles/s41467-019-12562-2}.

\bibitem[S68]{Skoblin.2018}
\bibinfo{author}{Skoblin, G.}, \bibinfo{author}{Sun, J.} \& \bibinfo{author}{Yurgens, A.}
\newblock \bibinfo{title}{{Graphene bolometer with thermoelectric readout and capacitive coupling to an antenna}}.
\newblock \emph{\bibinfo{journal}{Applied Physics Letters}} \textbf{\bibinfo{volume}{112}}, \bibinfo{pages}{063501} (\bibinfo{year}{2018}).
\newblock \urlprefix\url{http://aip.scitation.org/doi/10.1063/1.5009629}.

\bibitem[S69]{Yuan.2020}
\bibinfo{author}{Yuan, S.} \emph{et~al.}
\newblock \bibinfo{title}{{Room Temperature Graphene Mid-Infrared Bolometer with a Broad Operational Wavelength Range}}.
\newblock \emph{\bibinfo{journal}{ACS Photonics}} \textbf{\bibinfo{volume}{7}}, \bibinfo{pages}{1206--1215} (\bibinfo{year}{2020}).

\bibitem[S70]{Han.2013}
\bibinfo{author}{Han, Q.} \emph{et~al.}
\newblock \bibinfo{title}{{Highly sensitive hot electron bolometer based on disordered graphene}}.
\newblock \emph{\bibinfo{journal}{Scientific Reports}} \textbf{\bibinfo{volume}{3}}, \bibinfo{pages}{3533} (\bibinfo{year}{2013}).
\newblock \urlprefix\url{http://www.nature.com/srep/2013/131218/srep03533/full/srep03533.html}.

\bibitem[S71]{Hruby.Neugebauer.2024}
\bibinfo{author}{Hrubý, J.} \emph{et~al.}
\newblock \bibinfo{title}{{Graphene quantum dot bolometer for on-chip detection of organic radical}}.
\newblock \emph{\bibinfo{journal}{Applied Physics Letters}} \textbf{\bibinfo{volume}{124}}, \bibinfo{pages}{123505} (\bibinfo{year}{2024}).

\bibitem[S72]{Sassi.2017}
\bibinfo{author}{Sassi, U.} \emph{et~al.}
\newblock \bibinfo{title}{{Graphene-based mid-infrared room-temperature pyroelectric bolometers with ultrahigh temperature coefficient of resistance}}.
\newblock \emph{\bibinfo{journal}{Nature Communications}} \textbf{\bibinfo{volume}{8}}, \bibinfo{pages}{14311} (\bibinfo{year}{2017}).
\newblock \eprint{1608.00569}.

\bibitem[S73]{Battista.Efetov.2024}
\bibinfo{author}{Battista, G.~D.} \emph{et~al.}
\newblock \bibinfo{title}{{Infrared single-photon detection with superconducting magic-angle twisted bilayer graphene}}.
\newblock \emph{\bibinfo{journal}{Science Advances}} \textbf{\bibinfo{volume}{10}}, \bibinfo{pages}{eadp3725} (\bibinfo{year}{2024}).

\bibitem[S74]{Halbertal.2017}
\bibinfo{author}{Halbertal, D.} \emph{et~al.}
\newblock \bibinfo{title}{{Imaging resonant dissipation from individual atomic defects in graphene}}.
\newblock \emph{\bibinfo{journal}{Science}} \textbf{\bibinfo{volume}{358}}, \bibinfo{pages}{1303 -- 1306} (\bibinfo{year}{2017}).
\newblock \urlprefix\url{http://www.sciencemag.org/lookup/doi/10.1126/science.aan0877}.

\bibitem[S75]{Bistritzer.2009}
\bibinfo{author}{Bistritzer, R.} \& \bibinfo{author}{Macdonald, A.~H.}
\newblock \bibinfo{title}{{Electronic Cooling in Graphene}}.
\newblock \emph{\bibinfo{journal}{Physical Review Letters}} \textbf{\bibinfo{volume}{102}}, \bibinfo{pages}{206410} (\bibinfo{year}{2009}).
\newblock \urlprefix\url{https://link.aps.org/doi/10.1103/PhysRevLett.102.206410}.

\bibitem[S76]{Betz.2012}
\bibinfo{author}{Betz, A.~C.} \emph{et~al.}
\newblock \bibinfo{title}{{Hot Electron Cooling by Acoustic Phonons in Graphene}}.
\newblock \emph{\bibinfo{journal}{Physical Review Letters}} \textbf{\bibinfo{volume}{109}}, \bibinfo{pages}{056805} (\bibinfo{year}{2012}).
\newblock \eprint{1203.2753}.

\bibitem[S77]{McKitterick.2016}
\bibinfo{author}{McKitterick, C.~B.}, \bibinfo{author}{Prober, D.~E.} \& \bibinfo{author}{Rooks, M.~J.}
\newblock \bibinfo{title}{{Electron-phonon cooling in large monolayer graphene devices}}.
\newblock \emph{\bibinfo{journal}{Physical Review B}} \textbf{\bibinfo{volume}{93}}, \bibinfo{pages}{075410} (\bibinfo{year}{2016}).
\newblock \urlprefix\url{http://link.aps.org/doi/10.1103/PhysRevB.93.075410}.

\bibitem[S78]{Draelos.2019o4h}
\bibinfo{author}{Draelos, A.~W.} \emph{et~al.}
\newblock \bibinfo{title}{{Subkelvin lateral thermal transport in diffusive graphene}}.
\newblock \emph{\bibinfo{journal}{Physical Review B}} \textbf{\bibinfo{volume}{99}}, \bibinfo{pages}{125427} (\bibinfo{year}{2019}).
\newblock \eprint{1812.11711}.

\bibitem[S79]{Viljas.2010}
\bibinfo{author}{Viljas, J.~K.} \& \bibinfo{author}{Heikkila, T.~T.}
\newblock \bibinfo{title}{{Electron-phonon heat transfer in monolayer and bilayer graphene}}.
\newblock \emph{\bibinfo{journal}{Physical Review B}} \textbf{\bibinfo{volume}{81}}, \bibinfo{pages}{245404} (\bibinfo{year}{2010}).

\bibitem[S80]{Song.2012}
\bibinfo{author}{Song, J. C.~W.} \& \bibinfo{author}{Levitov, L.~S.}
\newblock \bibinfo{title}{{Energy-Driven Drag at Charge Neutrality in Graphene}}.
\newblock \emph{\bibinfo{journal}{Physical Review Letters}} \textbf{\bibinfo{volume}{109}}, \bibinfo{pages}{236602} (\bibinfo{year}{2012}).
\newblock \urlprefix\url{http://xxx.lanl.gov/abs/1205.5257}.
\newblock \eprint{1205.5257}.

\bibitem[S81]{Graham.2013}
\bibinfo{author}{Graham, M.~W.}, \bibinfo{author}{Shi, S.-F.}, \bibinfo{author}{Ralph, D.~C.}, \bibinfo{author}{Park, J.} \& \bibinfo{author}{McEuen, P.~L.}
\newblock \bibinfo{title}{{Photocurrent measurements of supercollision cooling in graphene}}.
\newblock \emph{\bibinfo{journal}{Nature Physics}} \textbf{\bibinfo{volume}{9}}, \bibinfo{pages}{103--108} (\bibinfo{year}{2013}).
\newblock \eprint{1207.1249}.

\bibitem[S82]{Chen.2012}
\bibinfo{author}{Chen, W.} \& \bibinfo{author}{Clerk, A.~A.}
\newblock \bibinfo{title}{{Electron-phonon mediated heat flow in disordered graphene}}.
\newblock \emph{\bibinfo{journal}{Physical Review B}} \textbf{\bibinfo{volume}{86}}, \bibinfo{pages}{670} (\bibinfo{year}{2012}).
\newblock \urlprefix\url{https://link.aps.org/doi/10.1103/PhysRevB.86.125443}.

\bibitem[S83]{Fong.2013}
\bibinfo{author}{Fong, K.~C.} \emph{et~al.}
\newblock \bibinfo{title}{{Measurement of the Electronic Thermal Conductance Channels and Heat Capacity of Graphene at Low Temperature}}.
\newblock \emph{\bibinfo{journal}{Physical Review X}} \textbf{\bibinfo{volume}{3}}, \bibinfo{pages}{041008} (\bibinfo{year}{2013}).
\newblock \urlprefix\url{https://link.aps.org/doi/10.1103/PhysRevX.3.041008}.

\bibitem[S84]{Betz.2013}
\bibinfo{author}{Betz, A.~C.} \emph{et~al.}
\newblock \bibinfo{title}{{Supercollision cooling in undoped graphene}}.
\newblock \emph{\bibinfo{journal}{Nature Physics}} \textbf{\bibinfo{volume}{9}}, \bibinfo{pages}{109--112} (\bibinfo{year}{2013}).
\newblock \eprint{1210.6894}.

\bibitem[S85]{Massicotte.2021}
\bibinfo{author}{Massicotte, M.}, \bibinfo{author}{Soavi, G.}, \bibinfo{author}{Principi, A.} \& \bibinfo{author}{Tielrooij, K.-J.}
\newblock \bibinfo{title}{{Hot carriers in graphene – fundamentals and applications}}.
\newblock \emph{\bibinfo{journal}{Nanoscale}} \textbf{\bibinfo{volume}{13}}, \bibinfo{pages}{8376--8411} (\bibinfo{year}{2021}).
\newblock \eprint{2105.08352}.

\bibitem[S86]{Aamir.2021}
\bibinfo{author}{Aamir, M.~A.} \emph{et~al.}
\newblock \bibinfo{title}{{Ultrasensitive Calorimetric Measurements of the Electronic Heat Capacity of Graphene}}.
\newblock \emph{\bibinfo{journal}{Nano Letters}}  (\bibinfo{year}{2021}).
\newblock \urlprefix\url{https://pubs.acs.org/doi/10.1021/acs.nanolett.1c01553}.

\bibitem[S87]{Draelos.2019}
\bibinfo{author}{Draelos, A.~W.} \emph{et~al.}
\newblock \bibinfo{title}{{Supercurrent Flow in Multiterminal Graphene Josephson Junctions}}.
\newblock \emph{\bibinfo{journal}{Nano Letters}} \textbf{\bibinfo{volume}{19}}, \bibinfo{pages}{1039 -- 1043} (\bibinfo{year}{2019}).
\newblock \urlprefix\url{http://pubs.acs.org/doi/10.1021/acs.nanolett.8b04330}.

\bibitem[S88]{Song.2013ykg}
\bibinfo{author}{Song, J. C.~W.}, \bibinfo{author}{Tielrooij, K.~J.}, \bibinfo{author}{Koppens, F. H.~L.} \& \bibinfo{author}{Levitov, L.~S.}
\newblock \bibinfo{title}{{Photoexcited carrier dynamics and impact-excitation cascade in graphene}}.
\newblock \emph{\bibinfo{journal}{Physical Review B}} \textbf{\bibinfo{volume}{87}}, \bibinfo{pages}{155429} (\bibinfo{year}{2013}).
\newblock \urlprefix\url{http://link.aps.org/doi/10.1103/PhysRevB.87.155429}.

\bibitem[S89]{Brida.2013}
\bibinfo{author}{Brida, D.} \emph{et~al.}
\newblock \bibinfo{title}{{Ultrafast collinear scattering and carrier multiplication in graphene}}.
\newblock \emph{\bibinfo{journal}{Nature Communications}} \textbf{\bibinfo{volume}{4}}, \bibinfo{pages}{1987} (\bibinfo{year}{2013}).
\newblock \urlprefix\url{http://www.nature.com/doifinder/10.1038/ncomms2987}.

\bibitem[S90]{Ruzicka.2010}
\bibinfo{author}{Ruzicka, B.~A.} \emph{et~al.}
\newblock \bibinfo{title}{{Hot carrier diffusion in graphene}}.
\newblock \emph{\bibinfo{journal}{Physical Review B}} \textbf{\bibinfo{volume}{82}}, \bibinfo{pages}{195414} (\bibinfo{year}{2010}).
\newblock \eprint{1005.3850}.

\end{thebibliography}

\bibliographystyle{nature}

\end{document}